\documentclass{elsart}

 \usepackage{graphicx}

\usepackage{amssymb}

\def\beq{\begin{equation}}
\def\eeq{\end{equation}}
\def\be{\begin{equation}}
\def\ee{\end{equation}}

\def\bea{\begin{eqnarray}}
\def\eea{\end{eqnarray}}
\def\ba{\begin{array}}                  
\def\ea{\end{array}}

\def\d{\delta}
\def\D{\Delta}

\def\g{\gamma}

\begin{document}

\begin{frontmatter}



\title{Lorentz violation at high energy: concepts, phenomena and
astrophysical constraints}

\author[UMD]{Ted Jacobson}\ead{jacobson@physics.umd.edu},
\author[SISSA]{Stefano Liberati}\ead{liberati@sissa.it},
\author[DAVIS]{David Mattingly}\ead{mattingly@physics.ucdavis.edu}
\address[UMD]{Department of Physics, University of Maryland, USA}
\address[SISSA]{International School for Advanced Studies and INFN, Trieste, Italy}
\address[DAVIS]{Department of Physics, University of California at Davis, USA}

\begin{abstract}
We consider here the possibility of quantum gravity induced
violation of Lorentz symmetry (LV). Even if suppressed by the
inverse Planck mass such LV can be tested by current experiments and
astrophysical observations. We review the effective field theory
approach to describing LV, the issue of naturalness, and many
phenomena characteristic of LV. We discuss some of the current
observational bounds on LV, focusing mostly on those from high
energy astrophysics in the QED sector at order $E/M_{\rm Planck}$.
In this context we present a number of new results which include the
explicit computation of rates of the most relevant LV processes, the
derivation of a new photon decay constraint, and modification of
previous constraints taking proper account of the helicity
dependence of the LV parameters implied by effective field theory.
\end{abstract}

\end{frontmatter}

\section{Introduction}
\label{sec:LV?}

The discovery of Lorentz symmetry was one of the great advances in
the history of physics. This symmetry has been confirmed to ever
greater precision, and it powerfully constrains theories in a way
that has proved instrumental in discovering new laws of physics.
Moreover the mathematical structure of the Lorentz group is
compellingly simple. It is natural to assume under these
circumstances that Lorentz invariance is a symmetry of nature up to
arbitrary boosts. Nevertheless, there are several reasons to
question exact Lorentz symmetry. {F}rom a purely logical point of
view, the most compelling reason is that an infinite volume of the
Lorentz group is (and will always be) experimentally untested since,
unlike the rotation group, the Lorentz group is non-compact. Why
should we assume that {\it exact} Lorentz invariance holds when this
hypothesis cannot even in principle be tested?

While non-compactness may be a logically compelling reason to
question Lorentz symmetry, it is by itself not very encouraging.
However, there are also several reasons to suspect that there will
be a failure of Lorentz symmetry at some energy or boost. One reason
is the ultraviolet divergences of quantum field theory, which are a
direct consequence of the assumption that the spectrum of field
degrees of freedom is boost invariant. Another reason comes from
quantum gravity. Profound difficulties associated with the ``problem
of time'' in quantum gravity~\cite{Isham,Kuchar} have suggested that
an underlying preferred time may be necessary to make sense of this
physics, and general arguments suggest radical departures from
standard spacetime symmetries at the Planck scale~\cite{garay}.
Aside from general issues of principle, specific hints of Lorentz
violation have come from tentative calculations in various
approaches to quantum gravity: string theory tensor
VEVs~\cite{KS89}, cosmologically varying
moduli~\cite{Damour:1994zq}, spacetime foam~\cite{GAC-Nat},
semiclassical spin-network calculations in Loop QG~\cite{GP,loopqg},
non-commutative
geometry~\cite{Hayakawa,Mocioiu:2000ip,Carroll:2001ws,Anisimov:2001zc}, some
brane-world backgrounds~\cite{Burgess:2002tb}, and condensed matter
analogues of ``emergent gravity"~\cite{ABH}.

None of the above reasons amount to a convincing argument that
Lorentz symmetry breaking is a feature of quantum gravity. However,
taken together they do motivate the effort to characterize possible
observable consequences of LV and to strengthen observational
bounds. Moreover, apart from any theoretical motivation, significant
improvement of the precision with which fundamental symmetries are
tested is always desirable.

The study of the possibility of Lorentz violation is not new,
although it has recently received more attention because of both the
theoretical ideas just mentioned and improvements in observational
sensitivity and reach that allow even Planck suppressed Lorentz
violation to be detected (see e.g.~\cite{Mattingly:2005re} for an
extensive review). A partial list of such ``windows on quantum
gravity" is
\begin{itemize}
\item sidereal variation of LV couplings as the lab moves
  with respect to a preferred frame or directions
  \item cosmological variation of couplings
\item cumulative effects: long baseline dispersion and vacuum birefringence
  (e.g.~of signals from gamma ray bursts, active galactic
  nuclei, pulsars, galaxies)
\item new threshold reactions (e.g.~photon decay, vacuum
\v{C}erenkov effect)
\item shifted existing threshold reactions (e.g.~photon annihilation from
  blazars, GZK reaction)
\item LV induced decays not characterized by a
threshold (e.g.~decay of a particle from one helicity to the other
or photon splitting)
\item maximum velocity (e.g.~synchrotron peak from supernova
remnants)
\item dynamical effects of LV background fields (e.g.
  gravitational coupling and additional wave modes)
\end{itemize}

The possibility of interesting constraints (or observations) of LV
despite Planck suppression arises in different ways for the
different types of observations. In the laboratory experiments
looking for sidereal variations, the enormous number of atoms allow
variations of a resonance frequency to be measured extremely
accurately. In the case of dispersion or birefringence, the enormous
propagation distances would allow a tiny effect to accumulate. In
the new or shifted threshold case, the creation of a particle with
mass $m$ would be strongly affected by a LV term when the momentum
becomes large enough for this term to be comparable to the mass term
in the dispersion relation. Finally, an upper bound to electron
group velocity, even if very near the speed of light, can severely
limit the frequency of synchrotron radiation. We shall discuss
examples of all these phenomena.

The purpose of the present paper is twofold. First, we aim to give
an introductory overview of some of the important issues involved in
the consideration of Lorentz violation, including history,
conceptual basis and problems, observable phenomena, and the current
best constraints in certain sectors. Second, we present a number of
new results, including computations of rates of certain LV processes
and the derivation of a new photon decay constraint. We also analyze
the modifications of previous constraints that are required when the
helicity dependence of the LV parameters is properly taken into
account. In order to make the paper most useful as an introduction
to LV we have placed much material, including most of the technical
details of the new results, in appendices.

The structure of this paper is as follows. The next section gives a
historical summary of LV research, while section
\ref{parametrization} introduces the framework for parameterizing
LV, together with the conceptual issues this raises. In section
\ref{sec:dim5} we focus on the phenomenology of dimension 5 LV in
QED, and section \ref{constraints} presents the current constraints
on such LV. Constraints on other sorts of LV are surveyed in section
\ref{other}, focusing on ultra-high energy cosmic rays, and we close
in section~\ref{conclusions} with a discussion of future prospects.
Appendix \ref{apsec:synch} presents the analysis behind the LV
synchrotron constraint, appendix \ref{apsec:threshconfig} derives
the LV threshold configuration results, and appendix
\ref{apsec:rates} includes the computation of rates and thresholds
for some LV processes.

\section{A brief history of some LV research}
\label{history}

We present at this point a brief historical overview of research
related to Lorentz violation, mentioning some influential work but
without trying to be complete. For a more complete review see
Ref.~\cite{Mattingly:2005re}.

The idea of cosmological variation of coupling constants goes back
at least to Milne and Dirac beginning in the
1930's~\cite{Barrow:1997qh}, and continues to be of interest today.
This would be associated with a form of LV since the spacelike
surfaces on which the couplings are assumed constant would define a
local preferred frame. It has recently been stimulated both by the
string theoretic expectation that there are moduli fields, and by
controversial observational evidence for variation of the fine
structure constant $\alpha$~\cite{alpha-obs}. A set of related ideas
goes under the generic name of ``varying speed of light
cosmologies'' (VSL), which includes numerous distinct
formulations~\cite{VSL,EllisG}. New models and observational
constraints continue to be discussed in the literature (see
e.g.~\cite{BLMV,Bertolami}).

Suggestions of possible LV in particle physics go back at least to
the 1960's, when several authors wrote on that idea~\cite{Dirac}.
The possibility of LV in a metric theory of gravity was explored
beginning at least as early as the 1970's with work of Nordtvedt and
Will~\cite{LVmetr}. Such theoretical ideas were pursued in the '70's
and '80's notably by Nielsen and several other authors on the
particle theory side~\cite{7080theory}, and by Gasperini~\cite{Gasp}
on the gravity side.  A number of observational limits were obtained
during this period~\cite{HauganAndWill}.

Towards the end of the 80's Kostelecky and Samuel~\cite{KS}
presented evidence for possible spontaneous LV in string theory, and
motivated by this explored LV effects in gravitation.  The role of
Lorentz invariance in the ``trans--Planckian puzzle" of black hole
redshifts and the Hawking effect was emphasized in the early
90's~\cite{TedUltra}. This led to study of the Hawking effect for
quantum fields with LV dispersion relations commenced by
Unruh~\cite{Unruh} and followed up by others. Early in the third
millennium this line of research led to work on the related question
of the possible imprint of trans--Planckian frequencies on the
primordial fluctuation spectrum~\cite{BM}.  Meanwhile the
consequences of LV for particle physics were being explored using LV
dispersion relations e.g. by Gonzalez-Mestres~\cite{GM}.

Four developments in the late nineties seem to have stimulated a
surge of interest in LV research. One was a systematic extension of
the standard model of particle physics incorporating all possible LV
in the renormalizable sector, developed by Colladay and
Kosteleck\'{y}~\cite{CK}.  That provided a framework for computing
in effective field theory the observable consequences for many
experiments and led to much experimental work setting limits on the
LV parameters in the Lagrangian~\cite{AKbook}. On the observational
side, the AGASA experiment reported ultra high energy (UHE) cosmic
ray events beyond the GZK proton cutoff~\cite{GZK,Takeda:1998ps}.
Coleman and Glashow then suggested the possibility that LV was the
culprit in the possibly missing GZK cutoff~\cite{CG-GZK}\footnote{
Remarkably, already in 1972 Kirzhnits and Chechin~\cite{Kirzhnits}
explored the
  possibility that an apparent missing cutoff in the UHE
  cosmic ray spectrum could be explained by something
  that looks very similar to the recently proposed
  ``doubly special relativity''~\cite{DSR}.},
and explored many other high energy consequences of renormalizable,
isotropic LV leading to different limiting speeds for different
particles~\cite{CGlong}. In the fourth development, it was pointed
out by Amelino-Camelia et al~\cite{GAC-Nat} that the sharp high
energy signals of gamma ray bursts could reveal LV photon dispersion
suppressed by one power of energy over the mass $M\sim 10^{-3}M_{\rm
P}$, tantalizingly close to the Planck mass.

Together with the improvements in observational reach mentioned
earlier, these developments attracted the attention of a large
number of researchers to the subject. Shortly after
Ref.~\cite{GAC-Nat} appeared, Gambini and Pullin~\cite{GP} argued
that semiclassical loop quantum gravity suggests just such
LV.\footnote{Some later work supported this notion, but the issue
continues to be debated~\cite{KP,Alfaro:2004ur}. In any case, the
dynamical aspect of the theory is not under enough control at this
time to make any definitive statements concerning LV.} Following
this work, a very strong constraint on photon birefringence was
obtained by Gleiser and Kozameh~\cite{GK} using UV light from
distant galaxies.
Further stimulus came from the suggestion~\cite{PM} that an LV
threshold shift might explain the apparent under-absorption on the
cosmic IR background of TeV gamma rays from the blazar Mkn501,
however it is now believed by many that this anomaly goes away when
a corrected IR background is used~\cite{Kono}.

The extension of the effective field theory (EFT) framework to
incorporate LV dispersion relations suppressed by the ratio
$E/M_{\rm Planck}$ was performed by Myers and Pospelov~\cite{MP}.
This allowed $E/M_{\rm Planck}$ LV to be explored in a systematic
way. The use of EFT also imposes certain relations between the LV
parameters for different helicities, which strengthened some prior
constraints while weakening others. Using EFT a very strong
constraint~\cite{Crab} on the possibility of a maximum electron
speed less than the speed of light was deduced from observations of
synchrotron radiation from the Crab Nebula. However, as discussed
here, this constraint is weakened due to the helicity and
particle/anti-particle dependence of the LV parameters

\section{Parametrization of Lorentz violation}
\label{parametrization}

A simple approach to a phenomenological description of LV is via
deformed dispersion relations. This approach was adopted in much
work on the subject, and it seems to afford a relatively
theory-independent framework in which to explore the unknown
possibilities. On the other hand, not much can really be predicted
with confidence just based on free particle dispersion relations,
without the use of both conservation laws and interaction dynamics.
Hence one is led to adopt a more comprehensive LV model in order to
deduce meaningful constraints. In this section we discuss these
ideas in turn, ending with a focus on the use of effective field
theory, which provides a well-motivated and unambiguous hypothesis
that can be tested.

\subsection{Deformed dispersion relations}
\label{subsec:disp}

If rotation invariance and analyticity around $p=0$ are assumed the
dispersion relation for a given particle type can be written as
\begin{equation}
E^2=p^2 + m^2 + \Delta(p), \label{gen-disprel}
\end{equation}
where $E$ is the energy, $p$ is hereafter the magnitude of the
three-momentum, and
\beq \Delta(p)= \tilde{\eta}_1 p^1 + \tilde{\eta}_2 p^2 +
\tilde{\eta}_3 p^3 + \tilde{\eta}_4 p^4 +\cdots \label{disprel1}
\eeq
Since this relation is not Lorentz invariant, the frame in which it
applies must be specified. Generally this is taken to be the average
cosmological rest frame, i.e. the rest frame of the cosmic microwave
background.~\footnote{There are attempts to interpret such deformed
dispersion relations as Casimir invariants of a new relativity group
which incorporates two invariant scales, $c$ and $M_{\rm Pl}$,
instead of just the speed of light like in Special Relativity (SR).
These theories are generally called Doubly (or Deformed) Special
Relativity (DSR)~\cite{DSR}.}

Let us introduce two mass scales, $M=10^{19}\, {\rm
  GeV}\approx M_{\rm Planck}$, the putative scale of
quantum gravity, and $\mu$, a particle physics mass scale.  To keep
mass dimensions explicit we factor out possibly appropriate powers
of these scales, defining the dimensionful $\tilde{\eta}$'s in terms
of corresponding dimensionless parameters. It might seem natural
that the $p^n$ term with $n\ge3$ be suppressed by $1/M^{n-2}$, and
indeed this has been assumed in many works. But following this
pattern one would expect the $n=2$ term to be unsuppressed and the
$n=1$ term to be even more important. Since any LV at low energies
must be small, such a pattern is untenable. Thus either there is a
symmetry or some other mechanism protecting the lower dimension
operators from large LV, or the suppression of the higher dimension
operators is greater than $1/M^{n-2}$.  This is an important issue
to which we return in section~\ref{naturalness}.

For the moment we simply follow the observational lead and insert at
least one inverse power of $M$ in each term, viz.
\beq \tilde{\eta}_1=\eta_1 \frac{\mu^2}{M},\qquad
\tilde{\eta}_2=\eta_2 \frac{\mu}{M},\qquad \tilde{\eta}_3=\eta_3
\frac{1}{M},\qquad \tilde{\eta}_4=\eta_4 \frac{1}{M^2}.
\label{disprel2} \eeq
In characterizing the strength of a constraint we refer to the
$\eta_n$ without the tilde, so we are comparing to what might be
expected from Planck-suppressed LV. We allow the LV parameters
$\eta_i$ to depend on the particle type, and indeed it turns out
that they {\it
  must} sometimes be different but related in certain
ways for photon polarization states, and for particle and
antiparticle states, if the framework of effective field theory is
adopted. In an even more general setting, Lehnert~\cite{Leh} studied
theoretical constraints on this type of LV and deduced the necessity
of some of these parameter relations.

The deformed dispersion relations are introduced for individual
particles only; those for macroscopic objects are then inferred by
addition. For example, if $N$ particles with momentum $\bf p$ and
mass $m$ are combined, the total energy, momentum and mass are
$E_{\rm tot}=NE(p)$, ${\bf p}_{\rm tot}=N{\bf p}$, and $m_{\rm
tot}=Nm$, so that $E_{\rm tot}^{2}=p_{\rm tot}^{2} + m_{\rm
tot}^{2}+ N^2\D(p)$. Although the Lorentz violating term can be
large in some fixed units, its ratio with the mass and momentum
squared terms in the dispersion relation is the same as for the
individual particles. Hence, there is no observational conflict with
standard dispersion relations for macroscopic objects.

This general framework allows for superluminal propagation, and
spacelike 4-momentum relative to a fixed background metric.  It has
been argued~\cite{KL} that this leads to problems with causality and
stability. In the setting of a LV theory with a single preferred
frame, however, we do not share this opinion. We cannot see any room
for such problems to arise, as long as in the preferred frame the
physics is guaranteed to be causal and the states all have positive
energy.

\subsection{The need for a more complete framework}

Various different theoretical approaches to LV have been taken. Some
researchers restrict attention to LV described in the framework of
effective field theory (EFT), while others allow for effects not
describable in this way, such as those that might be due to
stochastic fluctuations of a ``space-time foam''. Some restrict to
rotationally invariant LV, while others consider also rotational
symmetry breaking. Both true LV as well as ``deformed" Lorentz
symmetry (in the context of so-called ``doubly special
relativity"\cite{DSR}) have been pursued. Another difference in
approaches is whether or not different particle types are assumed to
have the same LV parameters.

The rest of this article will focus on just one of these approaches,
namely LV describable by standard EFT, assuming rotational
invariance, and allowing distinct LV parameters for different
particles.  This choice derives from the attitude that, in exploring
the possible phenomenology of new physics, it is useful to retain
enough standard physics so that clear predictions can be made, and
to keep the possibilities narrow enough to be meaningfully
constrained.  Furthermore, of the LV phenomena mentioned in the
Introduction, only dispersion and birefringence are determined
solely by the kinematic dispersion relations. Analysis of threshold
reactions obviously requires in addition an assumption of
energy-momentum conservation, and to impose constraints the reaction
rates must be known. This requires knowledge of matrix elements, and
hence the dynamics comes into play.  We therefore need a complete
(at least at low energy) theory to properly derive constraints. Only
EFT currently satisfies this requirement.

This approach is not universally favored (for an example of a
different approach see~\cite{GAC-crit}). Therefore we think it is
important to spell out the motivation for the choices we have made.
First, while of course it may be that EFT is not adequate for
describing the leading quantum gravity phenomenology effects, it has
proven very effective and flexible in the past. It produces local
energy and momentum conservation laws, and seems to require for its
applicability just locality and local spacetime translation
invariance above some length scale. It describes the standard model
and general relativity (which are presumably not fundamental
theories), a myriad of condensed matter systems at appropriate
length and energy scales, and even string theory (as perhaps most
impressively verified in the calculations of black hole entropy and
Hawking radiation rates).  It is true that, e.g., non-commutative
geometry (NCG) can lead to EFT with problematic IR/UV mixing,
however this more likely indicates a physically unacceptable feature
of such NCG rather than a physical limitation of EFT.

It is worth remarking that while we choose EFT so as to work in a
complete and well motivated framework, in fact many constraints are
actually quite insensitive to the specific dynamics of the theory,
other than that it obeys energy-momentum conservation. As an
example, consider photon decay. In ordinary Lorentz invariant
physics, photon decay into an electron/positron pair is forbidden
since two timelike four-momenta (the outgoing pair) cannot add up to
the null four momentum of the photon.  With LV the photon four
momentum can be timelike and therefore photons above a certain
energy can be unstable. Once the reaction is kinematically allowed,
the photon lifetime is extremely short ($\ll 1$ sec) when calculated
with standard QED plus modified dispersion, much shorter for example
than the required lifetime of $10^{11}$ seconds for high energy
photons that reach us from the Crab nebula. Thus we could tolerate
huge modifications to the matrix element (the dynamics) and still
have a photon decay rate incompatible with observation. Hence, often
the dynamics isn't particularly important when deriving constraints.

The assumption of rotational invariance is motivated by the idea
that LV may arise in QG from the presence of a short distance
cutoff. This suggests a breaking of boost invariance, with a
preferred rest frame\footnote{See however~\cite{Dowker:2003hb} for
an example where (coarse grained) boost invariance is preserved in a
discrete model, and~\cite{Livine:2004xy} for a study of how
discreteness may be compatible with Lorentz symmetry in a quantum
setting.} but not necessarily an observable breaking of rotational
invariance. Note also that it is very difficult to construct a
theory that breaks rotation invariance while preserving boost
invariance. (For example, if a spacelike four-vector is introduced
to break rotation invariance, the four-vector also breaks boost
invariance.) Since a constraint on pure boost violation is, barring
a conspiracy, also a constraint on boost plus rotation violation, it
is sensible to simplify with the assumption of rotation invariance
at this stage. The preferred frame is assumed to coincide with the
rest frame of the CMB since the Universe provides this and no other
candidate for a cosmic preferred frame.

Finally why do we choose to complicate matters by allowing for
different LV parameters for different particles? First, EFT for
first order Planck suppressed LV (see section~\ref{subsec:EFTLV})
requires this for different polarizations or spin states, so it is
unavoidable in that sense. Second, we see no reason {\it a priori}
to expect these parameters to coincide.  The term ``equivalence
principle'' has been used to motivate the equality of the
parameters. However, in the presence of LV dispersion relations,
free particles with different masses travel on different
trajectories even if they have the same LV
parameters~\cite{Fischbach:wq,Jacobson:2002hd}.  Moreover, different
particles would presumably interact differently with the spacetime
microstructure since they interact differently with themselves and
with each other. For an explicit example see~\cite{Alfaro:2005sf}.
Another example of this occurs in the braneworld model discussed in
Ref.~\cite{Burgess:2002tb}, and an extreme version occurs in the
proposal of Ref.~\cite{emn} in which only certain particles feel the
spacetime foam effects. (Note however that in this proposal the LV
parameters fluctuate even for a given kind of particle, so EFT would
not be a valid description.)

\subsection{Effective field theory and LV} \label{subsec:EFTLV}

In this subsection we briefly discuss some explicit formulations of
LVEFT. First, the (minimal) standard model extension (SME) of
Colladay and Kosteleck\'{y}~\cite{CK} consists of the standard model
of particle physics plus all Lorentz violating renormalizable
operators (i.e. of mass dimension $\le4$) that can be written
without changing the field content or violating the gauge symmetry.
For illustration, the leading order terms in the QED sector are the
dimension three terms
\beq -b_a\bar{\psi}\gamma_5 \gamma^a \psi
-\frac{1}{2}H_{ab}\bar{\psi}\sigma^{ab}\psi \eeq
and the dimension four terms
\beq -\frac{1}{4}k^{abcd}F_{ab}F_{cd} +\frac{i}{2}\bar{\psi}(c_{ab}+
d_{ab}\gamma_5)\gamma^a \stackrel{\leftrightarrow}{D}\!\!{}^b\psi,
\eeq
where the dimension one coefficients $b_a$, $H_{ab}$ and
dimensionless $k^{abcd}$, $c_{ab}$, and $d_{ab}$ are constant
tensors characterizing the LV.  If we assume rotational invariance
then these must all be constructed from a given unit timelike vector
$u^a$ and the Minkowski metric $\eta_{ab}$, hence $b_a\propto u_a$,
$H_{ab}=0$, $k^{abcd}\propto u^{[a} \eta^{b][c} u^{d]}$,
$c_{ab}\propto u_au_b$, and $d_{ab}\propto u_au_b$. Such LV is thus
characterized by just four numbers.

The study of Lorentz violating EFT in the higher mass dimension
sector was initiated by Myers and Pospelov~\cite{MP}. They
classified all LV dimension five operators that can be added to the
QED Lagrangian and are quadratic in the same fields, rotation
invariant, gauge invariant, not reducible to a combination of lower
and/or higher dimension operators using the field equations, and
contribute $p^3$ terms to the dispersion relation. Just three
operators arise:
\beq
-\frac{\xi}{2M}u^mF_{ma}(u\cdot\partial)(u_n\tilde{F}^{na})+\frac{1}{2M}
u^m\bar{\psi}\gamma_m(\zeta_1+\zeta_2\gamma_5)(u\cdot
\partial)^2\psi \label{dim5} \eeq
where $\tilde{F}$ denotes the dual of $F$, and $\xi,\zeta_{1,2}$ are
dimensionless parameters. The sign of the $\xi$ term in (\ref{dim5})
is opposite to that in~\cite{MP}, and is chosen so that positive
helicity photons have $+\xi$ for a dispersion coefficient (see
below). Also the factor 2 in the denominator is introduced to avoid
a factor of 2 in the dispersion relation for photons. All of the
terms (\ref{dim5}) violate CPT symmetry as well as Lorentz
invariance. Thus if CPT were preserved, these LV operators would be
forbidden.

\subsection{Naturalness of small LV at low energy?}
\label{naturalness}

As discussed above in subsection \ref{subsec:disp}, if LV operators
of dimension $n>4$ are suppressed as we have imagined by
$1/M^{n-2}$, LV would feed down to the lower dimension operators and
be strong at low energies~\cite{CGlong,MP,PerezSudarsky,Collins},
unless there is a symmetry or some other mechanism that protects the
lower dimension operators from strong LV. What symmetry (other than
Lorentz invariance, of course!) could that possibly be?

In the Euclidean context, a discrete subgroup of the Euclidean
rotation group suffices to protect the operators of dimension four
and less from violation of rotation symmetry. For
example~\cite{Hyper}, consider the ``kinetic'' term in the EFT for a
scalar field with hypercubic symmetry,
$M^{\mu\nu}\partial_\mu\phi\partial_\nu\phi$. The only tensor
$M^{\mu\nu}$ with hypercubic symmetry is proportional to the
Kronecker delta $\delta^{\mu\nu}$, so full rotational invariance is
an ``accidental'' symmetry of the kinetic operator.

If one tries to mimic this construction on a Minkowski lattice
admitting a discrete subgroup of the Lorentz group, one faces the
problem that each point has an infinite number of neighbors related
by the Lorentz boosts. For the action to share the discrete symmetry
each point would have to appear in infinitely many terms of the
discrete action, presumably rendering the equations of motion
meaningless.

Another symmetry that could do the trick is three dimensional
rotational symmetry together with a symmetry between different
particle types. For example, rotational symmetry would imply that
the kinetic term for a scalar field takes the form
$(\partial_t\phi)^2-c^2(\partial_i\phi)^2$, for some constant $c$.
Then, for multiple scalar fields, a symmetry relating the fields
would imply that the constant $c$ is the same for all, hence the
kinetic term would be Lorentz invariant with $c$ playing the role of
the speed of light.  Unfortunately this mechanism does not work in
nature, since there is no symmetry relating all the physical fields.

Perhaps under some conditions a partial symmetry could be adequate,
e.g. grand unified gauge and/or super symmetry. In fact, some recent
work~\cite{Nibbelink:2004za,Jain:2005as} presents evidence that
supersymmetry can indeed play this role. Supersymmetry (SUSY) here
refers to the symmetry algebra that is a `square root' of the
spacetime translation group. The nature of this square root depends
upon the Minkowski metric, so is tied to the Lorentz group, but it
does not require Lorentz symmetry. Nibbelink and Pospelov showed in
Ref.~\cite{Nibbelink:2004za}, using the superfield formalism, that
the SUSY and gauge symmetry preserving LV operators that can be
added to the SUSY Standard Model first appear at dimension five.
This solves the naturalness problem in the sense of having
Planck-suppressed dimension five operators without accompanying huge
lower dimension LV operators. However, it should be noted that in
this scenario the $O(p^3)$ terms in the particle dispersion
relations are suppressed by an additional factor of $m^2/M^2$
compared to (\ref{disprel1},\ref{disprel2}).

In a different analysis, Jain and Ralston~\cite{Jain:2005as} showed
that if the Wess-Zumino model is cut off at the energy scale
$\Lambda$ in a LV manner that preserves SUSY, then LV radiative
corrections to the scalar particle self-energy are suppressed by the
ratio $m^2/\Lambda^2$, unlike in the non-SUSY case~\cite{Collins}
where they diverge logarithmically with $\Lambda$. This is another
example of SUSY making approximate low energy Lorentz symmetry
natural in the presence of large high energy LV.

Of course SUSY is broken in the real world. LV SUSY QED with softly
broken SUSY was recently studied in~\cite{Bolokhov:2005cj}. It was
found there that, upon SUSY breaking, the dimension five SUSY
operators generate dimension three operators large enough that the
dimension five operators must be suppressed by a mass scale much
greater than $M_{Planck}$. In this sense, the naturalness problem is
{\it not} solved in this setting (although it is not as severe as
with no SUSY). If CPT symmetry is imposed however, then all the
dimension five operators are excluded. After soft SUSY breaking, the
dimension six LV operators generate dimension four LV operators that
are currently not experimentally excluded. Hence perhaps this latter
scenario can solve the naturalness problem. But again, the SUSY LV
operators considered in~\cite{Bolokhov:2005cj} do not give rise to
the type of dispersion corrections we consider, hence the
astrophysical bounds discussed here are not relevant to that SUSY
model.

At this stage we assume the existence of some realization of the
Lorentz symmetry breaking scheme upon which constraints are being
imposed. If none exists, then our parametrization (\ref{disprel2})
is misleading, since there should be more powers of $1/M$
suppressing the higher dimension terms. In that case, current
observational limits on those terms do not significantly constrain
the fundamental theory.

\section{Phenomenology of QED with dimension 5 Lorentz violation}
\label{sec:dim5}

We now discuss in general terms the new phenomena arising when the
extra dimension five operators in (\ref{dim5}) are added to the QED
lagrangian. This lays the groundwork for the specific constraints
discussed in the next section.

Free particles and tree level interactions can be analyzed without
specifying the underlying mechanism that (we assume) protects
approximate low energy Lorentz symmetry, hence we restrict attention
to these. In principle radiative loop corrections cannot be avoided,
but their treatment requires some commitment as to the specific
mechanism.

The appearance of higher time derivatives in the Lagrangian brings
with it, if treated exactly, an increase in the number of degrees of
freedom. In the EFT framework however it is natural to truncate the
theory via perturbative reduction to the degrees of freedom that
exist without the higher derivative terms (see e.g.~\cite{Cheng}).
Although not discussed explicitly, it is this truncation we have in
mind in what follows.

\subsection{Free particle states}
\subsubsection{Photons}
In Lorentz gauge, $\partial^\mu A_\mu=0$, the free field equation
 of motion for $A_\mu$ in the preferred frame from (\ref{dim5}) is
\begin{equation} \label{eq:eomphoton0}
\Box A_\alpha= \frac {\xi} {2M} \epsilon_{\alpha \beta \gamma
\delta} u^{\beta} (u \cdot \partial)^2 F^{\gamma \delta}.
\end{equation}
For a particle travelling in the $z$-direction, the $A_{0,3}$
equations are the usual $\Box A_{0,3}=0$ and hence the residual
gauge freedom can still be applied to set $A_0=A_3=0$, leaving two
transverse polarizations with dispersion~\cite{MP}
\begin{equation}
\omega_{\pm}^2 = k^2 \pm \frac{\xi}{M}k^3.
 \label{photondisp}
\end{equation}
The photon subscripts $\pm$ denote right and left polarization,
$\epsilon^{\mu}_\pm=\epsilon^\mu_x \pm i \epsilon^\mu_y$, which have
opposite dispersion corrections as a result of the CPT violation in
the Lagrangian.

\subsubsection{Fermions} \label{subsubsec:fermions}
We now solve the modified free field Dirac equation for the electron
and positron eigenspinors and find the corresponding dispersion
relations. We shall see that there exists a basis of energy
eigenstates that also have definite helicity ${\bf \hat{p}}\cdot{\bf
J}$, hence helicity remains a good quantum number in the presence of
the LV dimension five operators.

Beginning with the electron, we assume the eigenspinor has the form
$u_s(p) e^{-i p \cdot x}$ where $p$ is the 4-momentum vector,
assumed to have positive energy, and $s=\pm1$ denotes positive and
negative helicity.  The field equation from (\ref{dim5}) in the
chiral basis implies
\begin{equation} \label{eq:diraceqn}
  A_R u_R = m u_L,\qquad
  A_L  u_L =  m u_R
\end{equation}
where
\begin{equation}\label{ARL}
A_{R,L}=E \mp {\bf p}\cdot{\bf \sigma}-\eta_\pm E^2/2M, \label{A}
\end{equation}
with $\eta_\pm=2(\zeta_1\pm \zeta_2)$ (the upper signs refer to
$A_R$) and $\sigma$ are the usual Pauli spin matrices. If the
spinors are helicity eigenspinors we can replace ${\bf p}\cdot{\bf
\sigma}$ with simply $sp$.  The dispersion relation is then
$A_RA_L=m^2$, or
\beq \label{eq:fermiondisp} %
E^2=m^2+p^2 + \frac {\eta_+} {2M} E^2(E+sp) + \frac {\eta_-} {2M}
E^2(E-sp) - \frac {\eta_+ \eta_-} {4M^2} E^4 \eeq
Moreover, the spinors $u_R$ and $u_L$ are proportional to each
other, with ratio $u_L/u_R = \sqrt{A_R/A_L}$. The eigenspinors for
the electron can thus be taken as
\begin{equation}
u_s(p)=\left(%
\begin{array}{c}
 \sqrt{A_R}\, \chi_{s}(p) \\
  \sqrt{A_L}\, \chi_{s}(p) \\
\end{array}%
\right)
=\left(%
\begin{array}{c}
 \sqrt{E-sp-\frac{\eta_+E^2}{2M}} \chi_{s}(p) \\
  \sqrt{E+sp-\frac{\eta_-E^2}{2M}}\chi_{s}(p) \\
\end{array}%
\right) \label{spinors}
\end{equation}
where $\chi_s$ is the two-component eigenspinor of $\hat{\bf p}
\cdot {\bf \sigma}$. with eigenvalue $s$. Note that the dispersion
relation $A_RA_L=m^2$ implies that either $A_R$ and $A_L$ are both
positive, or they are both negative. The definition (\ref{ARL})
shows that when the energy $E$ is positive and much smaller than
$M$, at least one of the two is always positive, hence they are both
positive. Then from (\ref{eq:diraceqn}) we see that $u_R$ and $u_L$
are related by a positive real multiple, implying that the square
roots in (\ref{spinors}) have a common sign.

The normalization of this spinor is $u_s^\dagger(p) u_s (p)= A_R
+A_L = 2E - 2 \zeta_1 E^2/M$. If the usual  factor $(2E)^{-1/2}$ is
included in the momentum integral in the field operator, the spinors
should be normalized to $2E$ to ensure the canonical commutation
relations hold with the standard annihilation and creation operators
assumed. Since we work only at energies $E \ll M$, the correction is
small and may be neglected.

The constraints we shall discuss arise from processes in which the
energy satisfies $m\ll E\ll M$. Then to lowest order in $m$ and
$E^2/M$ the dispersion relation (\ref{eq:fermiondisp}) for positive
and negative helicity electrons takes the form~\cite{MP}
\begin{equation} \label{eq:dispfermionhighE}
E_{\pm}^2 =p^2+m^2 +\eta_{\pm} \frac{p^3} {M},
\end{equation}
where we have replaced $E$ by $p$ in the last term, which is valid
to lowest order. At $E\gg m$ the helicity states take the
approximate form
\begin{equation}
u_+(p)\simeq \sqrt{2E}
\left(%
\begin{array}{c}
 \frac{m}{2E} \chi^+(p) \\
  \chi^+(p) \\
\end{array}%
\right),\qquad u_-(p)\simeq \sqrt{2E}
\left(%
\begin{array}{c}
 \chi^-(p) \\
 \frac{m}{2E}\chi^-(p) \\
\end{array}%
\right). \label{spinorshighE}
\end{equation}
These are almost chiral, with mixing still controlled by the mass as
in the usual Lorentz invariant case.

To find the dispersion relation and spinor wavefunctions for
positrons we can use the hole interpretation. A positron of energy,
momentum, and spin angular momentum $(E,{\bf p},{\bf J})$
corresponds to the {\it absence} of an electron with $(-E,{-\bf
p},{-\bf J})$. This electron state has the same helicity as the
positron state, since $(-{\bf \hat{p}})\cdot({-\bf J})={\bf
\hat{p}}\cdot{\bf J}$. Hence the positron dispersion relation is
obtained by the replacement $(E,s)\rightarrow(-E,s)$ in
(\ref{eq:fermiondisp}). This is equivalent to the replacement
$\eta_\pm\rightarrow -\eta_\mp$~\cite{JLMS}, from which we conclude
that the LV parameters for positrons are related to those for
electrons by
\beq \label{eq:ep} \eta^{\rm positron}_\pm=-\eta^{\rm electron}_\mp.
\eeq
The spinor wavefunction of the $(-E,{-\bf p},{-\bf J})$ electron
state---which is also the wavefunction multiplying the positron
creation operator in the expansion of the fermion field
operator---has the opposite spin state compared to the $(E,{\bf
p},{\bf J})$ electron, so in place of $\chi_s$ one has $\chi_{-s}$.
Since the energy $-E$ is negative, both $A_R$ and $A_L$ are now
negative. Hence the positron eigenspinors take the form
\begin{equation}
v_s(p)=\left(%
\begin{array}{c}
 \sqrt{|A_R|} \chi_{-s}(p) \\
  -\sqrt{|A_L|}\chi_{-s}(p) \\
\end{array}%
\right)
=\left(%
\begin{array}{c}
 \sqrt{E+sp+\frac{\eta_+E^2}{2M}} \chi_{-s}(p) \\
  -\sqrt{E-sp+\frac{\eta_-E^2}{2M}}\chi_{-s}(p) \\
\end{array}%
\right) \label{pspinors}
\end{equation}
At high energy these take the approximate form
\begin{equation}
v^+(p)\simeq \sqrt{2E}
\left(%
\begin{array}{c}
 \chi^-(p) \\
 -\frac{m}{2E}\chi^-(p)\\
\end{array}
\right), \qquad v^-(p)\simeq \sqrt{2E}
\left(%
\begin{array}{c}
 \frac{m}{2E} \chi^+(p) \\
  -\chi^+(p) \\
\end{array}%
\right). \label{pspinorshighE}
\end{equation}
This completes our analysis of the free particle states.

\subsection{LV signatures: free particles}

\subsubsection{Vacuum dispersion and birefringence}
The photon dispersion relation (\ref{photondisp}) entails two free
particle phenomena that can be used to look for Lorentz violation
and to constrain the parameter $\xi$:

(i) The propagation speed depends upon both frequency and
polarization, which would produce dispersion in the {\it time of
flight} of different frequency components of a signal originating at
the same event.

(ii) Linear polarization direction is rotated through a
frequency-dependent angle, due to different phase velocities for
opposite helicities. This {\it vacuum birefringence} would rotate
the polarization direction of monochromatic radiation, or could
depolarize linearly polarized radiation composed of a spread of
frequencies.

\subsubsection{Limiting speed of charged particles}

The electron or positron dispersion relations
(\ref{eq:dispfermionhighE}) imply that if the $\eta$-parameter is
negative for a given helicity, that helicity has a maximum
propagation speed which is strictly less than the low frequency
speed of light. This LV phenomenon would limit the frequency of
synchrotron radiation produced by that helicity. This is not
strictly a free particle effect since it involves acceleration in an
external field and radiation, but the essential LV feature is the
limiting speed. In the Appendix section \ref{apsec:synch} we review
the EFT analysis of  this phenomenon.

\subsection{LV signatures: particle interactions}

There are a number of LV effects involving particle interactions
that do not occur in ordinary Lorentz invariant QED. These effects
include photon decay ($\gamma \rightarrow e^+e^-$), vacuum Cerenkov
radiation and helicity decay ($e^- \rightarrow e^- \gamma$),
fermion pair emission ($e^- \rightarrow e^- e^-e^+$),
and photon splitting ($\gamma\rightarrow n\gamma$).  Additionally, the
threshold for the ``photon absorption'' ($\gamma\gamma\rightarrow
e^+e^-$) is shifted away from its Lorentz invariant value.

All but photon splitting have nonvanishing tree level amplitudes,
hence can be considered without getting involved in the question of
the UV completion of the theory. Once loop amplitudes are
considered, the UV completion becomes important, and as discussed in
section \ref{naturalness} can produce large effects at low energy
unless there is fine tuning or a symmetry protection mechanism.

The threshold effects can occur at energies many orders of magnitude
below the Planck scale. To see why, note that thresholds are
determined by particle masses, hence if the $O(E/M)$ term in
Eqs.~(\ref{photondisp}) or (\ref{eq:dispfermionhighE}) is comparable
to the electron mass term, $m^2$, one can expect a significant
threshold shift. For LV coefficients of order one this occurs at the
momentum
\begin{equation}
   p_{\rm dev}\sim\left({m^2 M}\right)^{1/3}\approx 10\, {\rm TeV},
 \label{eq:equal}
\end{equation}
which gives a rough idea of the energies one needs to reach in order
to put constraints of order one on the Lorentz violations considered
here.~\footnote{See also~\cite{Jacobson:2002hd} for a generalization
to arbitrary order of Lorentz violation and different particle
sectors.}

To use the anomalous decay processes for constraining LV one needs
to know the threshold energy (if any) and how the rate depends on
the incoming particle energy and the LV parameters. The details of
computing these thresholds and rates are relegated to the Appendix
sections \ref{apsec:threshconfig} and \ref{apsec:rates}. Here we
just cite the main results.

\subsubsection{Photon decay}
\label{photondecay} A photon with energy above a certain threshold
can decay to an electron and a positron, $\gamma \rightarrow
e^+e^-$. The threshold and rate for this process generally depend on
all three LV parameters $\xi$ and $\eta_\pm$. Earlier use of the
photon decay constraint with the dispersion relation
(\ref{photondisp}) did not allow for helicity and
particle/anti-particle dependence of the LV parameters. To obtain
constraints on just two parameters, but consistent with EFT, we can
focus on processes in which only either $\eta_+$ or $\eta_-$ is
involved, namely reactions in which the positron has opposite
helicity to the electron. For example, if the electron has positive
helicity then its LV parameter is $\eta_+$, and according to
(\ref{eq:ep}) the LV parameter for the negative helicity positron is
$-\eta_+$.

At threshold, the final particle momenta are all aligned (cf.
Appendix~\ref{apsec:threshconfig}). Since the incoming photon has
nonzero spin, angular momentum cannot be conserved if the electron
and positron have opposite helicities. However, the momenta {\it
above} threshold need not be aligned, so that angular momentum can
be conserved with opposite helicities at any energy above threshold.
In this case, the rate vanishes at threshold and is suppressed very
close to the threshold, but above the threshold the rate quickly
begins to scale as
 $E^2/M$, where $E$
is the photon energy.\footnote{This corrects a prior
assertion~\cite{Jacobson:2001tu,Jacobson:2002hd} that the rate
scales as $E$.} For a 10 TeV photon, this corresponds to a decay
time of order $10^{-8}$  seconds (where we have  taken into account
the extra suppression coming from overall numerical factors, see
section~\ref{subsubsec:photondecayrate} in the Appendix). The
rapidity of the reaction (in comparison to the required lifetime of
$10^{11}$ seconds for an observed Crab photon) implies that a
threshold constraint, i.e. a bound on LV coefficients such that the
reaction does not happen at all, will be extremely close to the
constraint derived by requiring the lifetime to be longer than the
observed travel time.

\subsubsection{Vacuum \v{C}erenkov radiation and helicity decay}

The process $e^-\rightarrow e^-\gamma$ can either preserve or flip
the helicity of the electron. If the electron helicity is unchanged
we call it vacuum \v{C}erenkov radiation, while if the helicity
changes we call it helicity decay. Vacuum \v{C}erenkov radiation
occurs only above a certain energy threshold, and above that energy
the rate of energy loss scales as $p^3/M$, which implies that a 10
TeV electron would emit a significant fraction of its energy in
$10^{-9}$ seconds. Above the threshold energy, constraints derived
simply from threshold analysis alone are again reliable.

In contrast, if the positive and negative helicity LV parameters for
electrons are unequal, say $\eta_->\eta_+$, then decay from negative
to positive helicity can happen at any energy, i.e. there is no
threshold.\footnote{The lack of a threshold can easily be seen by
noting that the dispersion curves $E(p)$ of the two helicity
electrons can be connected by a null vector for any energy.
Therefore there is always a way to conserve energy-momentum with
photon emission (the photon four-vector is almost null even with
LV).} However it can be shown  (see Appendix \ref{subsec:HelDec})
that assuming generic order unity values of the LV parameters, the
rate is extremely small until the energy is comparable to the
threshold for \v{C}erenkov radiation. Above this energy, the rate is
$\sim e^2 m^2/p$ independent of the LV parameters, which (taking
into account all the numerical factors) implies that a 10 TeV
electron would flip helicity in about $10^{-9}$ seconds. Thus, above
this ``effective threshold", constraints on helicity decay derived
from the value of the effective threshold alone are reliable.

\subsubsection{Fermion pair emission}

The process $e^- \rightarrow e^-e^-e^+$ is similar to vacuum
\v{C}erenkov radiation or helicity decay, with the final photon
replaced by an electron-positron pair.  This reaction has been
studied previously in the context of the SME~\cite{KL,andrianov}.
Various combinations of helicities for the different fermions can be considered
individually. If we choose the particularly simple case (and the
only one we shall consider here) where all electrons have the same
helicity and the positron has the opposite helicity, then
according to Eq.~\ref{eq:ep} the threshold energy will depend on
only one LV parameter. In Appendix \ref{app:pairemission} we
derive the threshold for this reaction, finding that it is a
factor $\sim 2.5$ times higher than that for soft vacuum
\v{C}erenkov radiation. The rate for the reaction is high as well,
hence constraints may be imposed using just the value of the
threshold.

\subsubsection{Photon splitting}

The photon splitting processes $\gamma \rightarrow 2 \gamma$ and
$\gamma \rightarrow 3 \gamma$, etc.\ do not occur in standard QED.
Although there are corresponding Feynman diagrams (the triangle and
box diagrams), their amplitudes vanish.  In the presence of Lorentz
violation these processes are generally allowed when $\xi>0$.
However, the effectiveness of this reaction in providing constraints
depends heavily on the decay rate.

Aspects of vacuum photon splitting have been examined
in~\cite{Jacobson:2002hd,othersplitting}. An estimate of the rate,
independent of the particular form of the Lorentz violating theory
and neglecting the polarization dependence of the photon dispersion
relation, was given in Ref.~\cite{Jacobson:2002hd}. It was argued
that a lower bound on the lifetime is $\d^{-4}E^{-1}$, where $\d$ is
a Lorentz violating factor which for LV from the Myers-Pospelov
lagrangian is $\d\sim\xi E/M$. 
It was implicitly assumed in this argument that no large 
dimensionless ratios occur in the rate.
However, a recent paper~\cite{Gelmini:2005gy} 
shows that such a large ratio can indeed occur.

Ref.~\cite{Gelmini:2005gy} 
analyzes the case where 
there is no Lorentz violation 
in the electron-positron sector.
Then the (Lorentz-invariant) one-loop Euler-Heisenberg Lagrangian 
characterizes the photon interaction, even in the 
``rest frame" of the incoming photon. Using this interaction
it is argued that the lifetime is much shorter, by a
factor $(m_e/E_\gamma)^8\d^{-1}$, than the lower bound mentioned
in the previous paragraph. The extra factor of $\delta$ in the rate is compatible with the analysis of~\cite{Jacobson:2002hd} since there only the {\it minimum} number of
factors of $\delta$ was determined. However, 
the possibile appearance of
a large dimensionless factor like $(E_\gamma/m_e)^8$ in the rate was
overlooked in~\cite{Jacobson:2002hd}.

Using 50 TeV photons from the Crab nebula, about $10^{13}$ seconds away, the analysis of Ref.~\cite{Jacobson:2002hd}  
would have implied that the constraint on $\xi$ can be no stronger than $\xi\lesssim 10^4$, and even this is not competitive with the other constraints. However, for a 50 TeV photon the extra factor in the lifetime from the analysis of Ref.~\cite{Gelmini:2005gy} is of order $10^{-50}$, which would produce a bound $\xi\lesssim 10^{-3}$. Note
however that this analysis so far neglects the helicity dependence
of the photon dispersion, and assumes no Lorentz violation in
the electron-positron sector. 

\subsubsection{Photon absorption}

  A process related to photon decay is
photon absorption, $\g\g\rightarrow e^+e^-$. Unlike photon decay,
this is allowed in Lorentz invariant QED. If one of the photons has
energy $\omega_0$, the threshold for the reaction occurs in a
head-on collision with the second photon having the momentum
(equivalently energy) $k_{\rm LI}=m^2/\omega_{0}$. For $k_{\rm LI}=
10$ TeV (which is relevant for the observational constraints) the
soft photon threshold $\omega_0$ is approximately 25 meV,
corresponding to a wavelength of 50 microns.

In the presence of Lorentz violating dispersion relations the
threshold for this process is in general altered, and the process
can even be forbidden. Moreover, as noticed by
Klu\'zniak~\cite{Kluzniak}, in some cases there is an upper
threshold beyond which the process does not occur.\footnote{
A detailed investigation of upper thresholds was carried out by the
present authors in~\cite{Jacobson:2002hd,Mattingly:2002ba}. Our
results agree with those of \cite{Kluzniak} only in certain limiting
cases.}
The lower and upper thresholds for photon annihilation as a function
of the two parameters $\xi$ and $\eta$, were obtained
in~\cite{Jacobson:2002hd}, before the helicity dependence required
by EFT was appreciated. As the soft photon energy is low enough that
its LV can be ignored, this corresponds to the case where electrons
and positrons have the same LV terms. The  analysis is rather
complicated. In particular it is necessary to sort out whether the
thresholds are lower or upper ones, and whether they occur with the
same or different pair momenta.

The photon absorption constraint, neglecting helicity dependent
effects, came from the fact that LV can shift the standard QED
threshold for annihilation of multi--TeV $\g$-rays from nearby
blazars, such as Mkn 501, with the ambient infrared extragalactic
photons~\cite{Kluzniak,GAC-Pir,SG,Jacobson:2002hd,KonMaj,Jacobson:2003ty,S03}.
LV depresses the rate of absorption of one photon helicity, and
increases it for the other.  Although the polarization of the
$\g$-rays is not measured, the possibility that one of the
polarizations is essentially unabsorbed appears to be ruled out by
the observations which show the predicted attenuation~\cite{S03}.
The threshold analysis has not been redone to allow for the helicity
and particle/anti-particle dependence of $\eta$. The motivation for
doing so is not great since the constraint would at best not be
competitive with other constraints, and the power of the constraint
is limited by our ignorance of the source spectrum of emitted gamma
rays.

\section{Constraints on LV in QED at $O(E/M)$ }
\label{constraints}

In this section we discuss the  current observational constraints on
the LV parameters $\xi$ and $\eta_\pm$ in the Myers-Pospelov
extension of QED, focusing on those coming from high energy
processes. The highest observed energies occur in astrophysical
settings, hence it is from such observations that the strongest
constraints derive.

\begin{figure}[htb]
\vbox{ \hfil \scalebox{0.80}{{\includegraphics{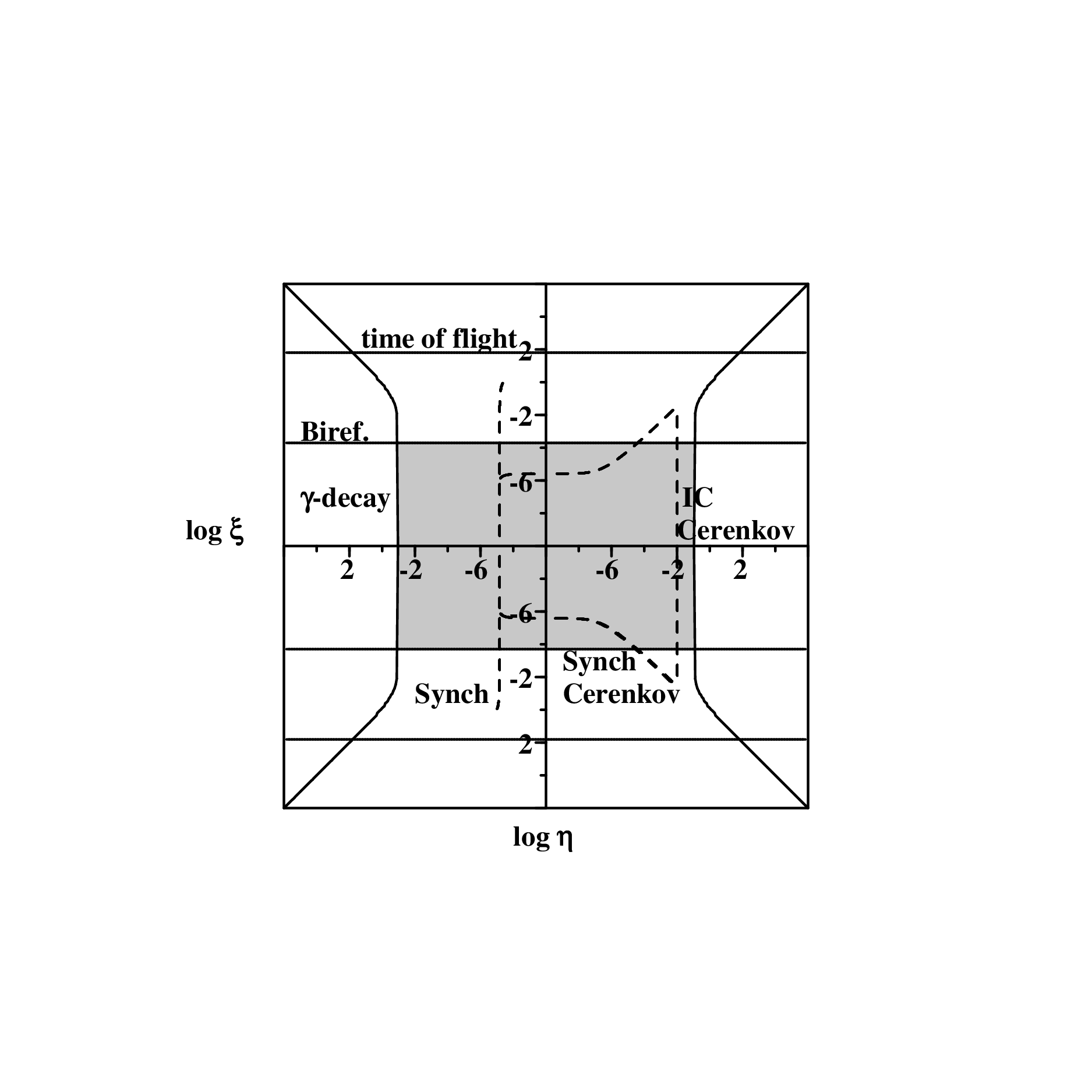}}} \hfil }
\bigskip
\caption{Constraints on LV in QED at $O(E/M)$ on a log-log plot. For
negative parameters minus the logarithm of the absolute value is
plotted, and region of width $10^{-10}$ is excised around each axis.
The constraints in solid lines apply to $\xi$ and both $\eta_\pm$,
and are symmetric about both the $\xi$ and the $\eta$ axis. At least
one of the two pairs $(\eta_\pm,\xi)$ must lie within the union of
the dashed bell-shaped region and its reflection about the $\xi$
axis. The IC and synchrotron \v{C}erenkov lines are truncated where
they cross.}
\label{fig:1}       
\end{figure}

The currently most useful constraints are summarized in
Fig.~\ref{fig:1}.
The allowed parameter space corresponds to the dark region of the
Figure. These constraints strongly bound Lorentz violation at order
$O(E/M)$ in QED, assuming the framework of effective field theory
holds. While the natural magnitude of the photon and electron
coefficients $\xi,\eta$ would be of order unity if there is one
power of suppression by the inverse Planck mass, the coefficients
are now restricted to the region $|\xi|\lesssim10^{-4}$ by
birefringence and $|\eta_{\pm}|\lesssim10^{-1}$ by photon decay.
The narrower bell-shaped region bounded by the dashed lines is
determined by the combination of synchrotron and \v{C}erenkov
constraints and applies only to one of the four pairs
$(\pm\eta_\pm,\xi)$. Equivalently, the union of this bell-shaped
region with its reflection about the $\xi$ axis applies to one of
the two pairs $(\eta_\pm,\xi)$. We shall now briefly discuss how
each constraint is obtained, leaving the details for the Appendices.

\subsection{Photon time of flight}
\label{sec:tof}

Photon time of flight constraints~\cite{TOF} limit differences
in the arrival time at Earth for photons originating in a distant
event~\cite{Pavlopoulos,GAC-Nat}. Time of flight can vary with
energy since the LV term in the group velocity is $\xi k/M$. The
arrival time difference for wave-vectors $k_1$ and $k_2$ is thus
\beq \Delta t=\xi (k_2-k_1) d/M,
  \label{timediff}
\eeq
which is proportional to the energy difference and the distance
travelled.  Constraints were obtained using the high energy radiation emitted by some gamma ray bursts (GRB) and active galaxies of the blazars class. The strength of such constraints is typically $\xi\lesssim O(100)$ or weaker~\cite{TOF}.

A possible problem with the above bounds is that in an emission event it is not known if the photons of different energies are produced
simultaneously. If different energies are emitted at different times
that might mask a LV signal. One way around this is to look for
correlations between time delay and redshift, which has been done
for a set of GRB's in~\cite{Ellis:2002in}. Since time of flight
delay is a propagation effect that increases over time, a survey of
GRB's at different redshifts can separate this from intrinsic source
effects. This enables constraints to be imposed (or LV to be
observed) despite uncertainty regarding source effects. While bounds
derived in this way are, at present, weaker than the bound presented
here, they are also more robust. One might also think that the
source uncertainties in GRB's could be mitigated by looking at high
energy narrow bursts, which by definition have nearly simultaneous
emission. However, for high energy narrow bursts the number of
photons per unit time can be very low, thereby limiting the shortest
detectable time lag~\cite{Piran:2004qe}.

The simultaneity problem may also be avoided if one uses the EFT
dispersion relation (\ref{photondisp}) for photons of the same
energy. Then one
can consider the velocity difference of the two polarizations at a
\textit{single} energy~\cite{JLMS}
\beq \Delta t=2|\xi|k/M.
  \label{timediff2}
\eeq
In principle this leads to a constraint at least twice as large as the one
arising from energy differences which is also independent of any intrinsic time lag between different energy photons. [It is possible, of course, that there is an
intrinsic helicity source effect, however this seems unlikely.]  In Fig.~\ref{fig:1} we use the
EFT improvement of the constraint of Biller et al.~\cite{TOF}, obtained using the blazar Markarian 421, an object whose distance is reliably known,
which yields $|\xi|<63$~\cite{JLMS}.  Note however that this
bound assumes that both polarizations are observed.  If for some
reason (such as photon decay of one polarization) only one
polarization is observed, then the bound shown in is Fig.~\ref{fig:1} weakened by a factor
of two. In any case the time of flight constraint remains many orders of magnitude weaker than the birefringence constraint so is rather irrelevant from an
EFT perspective.


\subsection{Birefringence}

The birefringence constraint arises from the fact that the LV
parameters for left and right circular polarized photons are
opposite (\ref{photondisp}).  The phase velocity thus depends on
both the wavevector and the helicity. Linear polarization is
therefore rotated through an energy dependent angle as a signal
propagates, which depolarizes an initially linearly polarized signal
comprised of a range of wavevectors. Hence the observation of
linearly polarized radiation coming from far away can constrain the
magnitude of the LV parameter.

In more detail, with the dispersion relation (\ref{photondisp}) the
direction of linear polarization is rotated through the angle
\beq \theta(t)=\left[\omega_+(k)-\omega_-(k)\right]t/2=\xi k^2 t/2M
\label{rotation} \eeq
for a plane wave with wave-vector $k$ over a propagation time $t$.
The difference in rotation angles for wave-vectors $k_1$ and $k_2$
is thus
\beq \Delta\theta=\xi (k_2^2-k_1^2) d/2M,
  \label{diffrotation}
\eeq
where we have replaced the time $t$ by the distance $d$ from the
source to the detector (divided by the speed of light). Note that
the effect is {\it quadratic} in the photon energy, and proportional
to the distance travelled. The constraint arises from the fact that
once the angle of polarization rotation differs by more than $\pi/2$
over the range of energies in a signal, the net polarization is
suppressed.

This effect has been used to constrain LV in the dimension three
(Chern-Simons)~\cite{CFJ}, four~\cite{KM} and five terms. The
strongest currently reliable constraint on the dimension five term
was deduced by Gleiser and Kozameh~\cite{GK} using UV light from
distant galaxies, and is given by $|\xi|\lesssim 2\times10^{-4}$.
The much stronger constraint $|\xi|\lesssim 2\times10^{-15}$ was
derived~\cite{JLMS,Mitro} from the report~\cite{CB03} of a high
degree of polarization of MeV photons from GRB021206. However, the
data has been reanalyzed in two different studies and no
statistically significant polarization was found~\cite{RF03}.

\subsection{The Crab nebula}

Apart from the two constraints just discussed, all the others
reported in Fig.~\ref{fig:1} arise from observations of the Crab
nebula. This object is the remnant  of a supernova that was observed
in 1054 A.D., and lies only about 1.9 Kpc from Earth. It is
characterized by the most energetic QED processes observed (e.g. it
is the source of the highest energy gamma rays) and is very well
studied. In contrast to the distant sources desirable for
constraints based on the cumulative effects of Lorentz violation,
the Crab nebula is nearby. This proximity facilitates the detection
of the low fluxes characteristic of the emission at the highest
energies, which is particularly useful for the remaining
constraints. Before undertaking a discussion of the constraints we
first summarize the nature of the Crab emission.

The Crab nebula is a bright source of radio, optical, X-ray and
gamma-ray emission. It exhibits a broad spectrum characterized by
two marked humps. This spectrum is consistently explained by a
combination of synchrotron emission by a high energy wind of
electrons and positrons, and inverse Compton scattering of the
synchrotron photons (plus perhaps 10\% other ambient photons) by the
same charges~\cite{AA96,deJager}. No other model for the emission is
under consideration, other than for producing some fraction of the
highest energy photons. For the constraints discussed here we assume
that this standard synchrotron/inverse Compton model is correct. In
contrast to our previous work, here we take fully into account the
role of the positrons, as well as the helicity and
particle/anti-particle dependence of the LV parameters. This
complicates matters and changes the nature of the constraints,
weakening some aspects and strengthening others.

The inverse Compton gamma ray spectrum of the Crab extends up to
energies of at least  50 TeV. The synchrotron emission has been
observed to extend at least up to energies of about 100
MeV~\cite{AA96}, just before the inverse Compton hump begins to
contribute to the spectrum. In standard Lorentz invariant QED, 100
MeV synchrotron radiation in a magnetic field of 0.6 mG would be
produced by electrons (or positrons) of energy 1500 TeV. The
magnetic field in the emission region has been estimated by several
methods which agree on a value between 0.15--0.6 mG (see e.g.
\cite{Hillas} and references therein). Two of these methods, radio
synchrotron emission and equipartition of energy, are insensitive to
Planck suppressed Lorentz violation, hence we are justified in
adopting a value of this order for the purpose of constraining
Lorentz violation. We use the largest value 0.6 mG for $B$, since it
yields the weakest constraint for the synchrotron radiation.

\subsection{Photon decay}

The observation of 50 TeV gamma rays emitted from the Crab nebula
implies that the threshold for photon decay
for at least one helicity
must be above 50 TeV. By
considering decays in which the electron and positron have opposite
helicity we can separately constrain both
$(\eta_+,\xi)$ and $(\eta_-,\xi)$
(cf. section~\ref{photondecay} and Appendix~\ref{apsec:rates}).
The allowed region is the union of those for which positive and
negative helicity photons do not decay.
This yields the constraint shown in Figure \ref{fig:1}.
The complete form of the allowed region was determined numerically.  However in
the region  $|\xi|< 10^{-4}$ defined by the birefringence
constraint, $\xi$ can be neglected compared to $\eta_\pm$, in which
case the constraint takes the analytic form
$|\eta_\pm|<6\sqrt{3}m^2M/k_{th}^3$. With $k_{th}=50\, {\rm Tev}$
this evaluates to $|\eta_\pm|\lesssim 0.2$.

\subsection{Vacuum \v{C}erenkov---Inverse Compton electrons}

The inverse Compton (IC) \v{C}erenkov constraint uses the electrons
and positrons of energy up to 50 TeV inferred via the observation of
50 TeV gamma rays from the Crab nebula which are explained by IC
scattering. Since the vacuum \v{C}erenkov rate is orders of
magnitude higher than the IC scattering rate, that process must not
occur for these charges~\cite{CGlong,Jacobson:2002hd} for at least
one of the four charge species (plus or minus helicity electron or
positron), and for either photon helicity. The excluded region in
parameter space is thus symmetric about the $\eta$-axis, but applies
only to one of the four parameters $\pm\eta_\pm$. That is, there is
a constraint that must be satisfied by either the pair
$(|\eta_+|,|\xi|)$ or the pair $(|\eta_-|,|\xi|)$.

The absence of the soft \v{C}erenkov threshold up to 50 TeV produces
the dashed vertical IC \v{C}erenkov line in Fig.~\ref{fig:1} (see
Appendix \ref{sebsec:cheren-thr} for a more detailed discussion of
the threshold). One can see from (\ref{pcerenkov}) that this yields
a constraint on $\eta$ of order $(11~{\rm TeV}/50~{\rm TeV})^3\sim
10^{-2}$. By itself this imposes no constraint at all on the
parameters $\eta_\pm$, since one of the two parameters $\pm\eta_+$
always satisfies it, as does one of $\pm\eta_-$. However, this is
not the whole story.

For parameters satisfying both $\xi<-3\eta$ and $\xi<\eta$ the
\v{C}erenkov threshold is ``hard", since it involves emission of a
high energy photon~\cite{Jacobson:2002hd,KonMaj}. This constraint
depends upon both $\xi$ and $\eta$, and is a curve on the parameter
space. We do not indicate this constraint in Fig.~\ref{fig:1} since
it is superseded by the synchrotron--hard \v{C}erenkov constraint
discussed below. It imposes a {\it lower} bound on one of
$|\eta_\pm|$ once $|\xi|$ is large enough.

\subsection{Fermion pair emission}

If we knew that all electrons and positrons were stable at 50 TeV,
then the threshold for all of them to undergo fermion pair
emission would necessarily be over 50 TeV. Using the processes
$e^- \rightarrow e^-e^-e^+$ and $e^+ \rightarrow e^+e^+e^-$ with
the helicities chosen so that only one of $\eta_\pm$ is involved
in the reaction, electron/positron stability would lead to the
constraint $|\eta_\pm|\lesssim 0.16$. This is a factor of 16
higher than that from the soft \v{C}erenkov threshold, and roughly
the same as the photon decay constraint. However, at least without
further detailed analysis of the Crab spectrum, all we can say is
that at least {\it one} of the fermion types is able to produce
the 50 TeV IC radiation. By itself this imposes no constraint at
all, since for each helicity either the electron or the positron
of the opposite helicity is stable to this particular pair
emission process.

\subsection{Synchrotron radiation}

An electron or positron with a negative value of $\eta$ has a
maximal group velocity less than the low energy speed of light,
hence there is a maximal synchrotron frequency $\omega_c^{\rm max}$
(\ref{eq:opeaklv2}) that it can produce, regardless of its
energy~\cite{Crab} (for details see Appendix~\ref{apsec:synch}).
Thus for at least one electron or positron helicity $\omega_c^{\rm
max}$ must be greater than the maximum observed synchrotron emission
frequency $\omega_{\rm obs}$. This yields the constraint
\begin{equation}
\eta > - \frac{M}{m}\left(\frac{0.34\, eB}{m\omega_{\rm
obs}}\right)^{3/2}. \label{eq:synchcon}
\end{equation}
The strongest constraint is obtained in the case of a system that
has the smallest $B/\omega_{\rm obs}$ ratio. This occurs in the Crab
nebula, which emits synchrotron radiation up to 100 MeV and has a
magnetic field no larger than 0.6 mG in the emitting region. Thus we
infer that at least one of $\pm\eta_\pm$ must be greater than
$-7\times10^{-8}$. This constraint is shown as a dashed line in
Fig.~\ref{fig:1}. As with the soft \v{C}erenkov constraint, by
itself this imposes no constraint on the parameters, but as
discussed in the next subsection it plays a role in a combined
constraint using also the \v{C}erenkov effect.

\subsection{Vacuum \v{C}erenkov effect---synchrotron electrons}
\label{subsec:synchcerenk}

The existence of the synchrotron producing charges can be exploited
to extend the vacuum \v{C}erenkov constraint. For a given $\eta$
satisfying the synchrotron bound, some definite electron energy
$E_{\rm synch}(\eta)$ must be present to produce the observed
synchrotron radiation. (As explained in Appendix~\ref{apsec:synch},
this is higher for negative $\eta$ and lower for positive $\eta$
than the Lorentz invariant value~\cite{Crab}.) Values of $|\xi|$ for
which the vacuum \v{C}erenkov threshold is lower than $E_{\rm
synch}(\eta)$ for either photon helicity can therefore be
excluded~\cite{JLMS}. This is always a hard photon threshold, since
the soft photon threshold occurs when the electron group velocity
reaches the low energy speed of light, whereas the velocity required
to produce any finite synchrotron frequency is smaller than this.
The corresponding constraint is shown by the dashed line labelled
``Synch. \v{C}erenkov" in Fig.~\ref{fig:1}. This constraint improves
on the current birefringence limit on $\xi$ in the parameter region
$\eta\lesssim 10^{-4}$. At least one of the four pairs
$(\pm\eta_\pm,|\xi|)$ must satisfy {\it both} this constraint and
the synchrotron constraint discussed above. This amounts to saying
that one of the two pairs $(|\eta_\pm|,|\xi|)$ must satisfy this
combined constraint {\it or} its reflection about the $\xi$ axis. As
with the hard \v{C}erenkov constraint discussed above, this imposes
a {\it lower} bound on one of $|\eta_\pm|$ once $|\xi|$ is large
enough.

The synchrotron and \v{C}erenkov constraints can further be linked,
to obtain an {\it upper} bound on one of $|\eta_\pm|$  as follows.
We know that (automatically) at least one of the four parameters
$\pm\eta_\pm$ satisfies the synchrotron constraint (a negative lower
bound), and at least one of them (automatically) satisfies the IC
\v{C}erenkov constraint (a positive upper bound). In fact, at least
one of the four parameters must satisfy {\it both} of these
constraints. The reason is that otherwise the synchrotron charges
would violate the \v{C}erenkov constraint, hence their energy would
necessarily be under 50 TeV, which is thirty times lower than the
Lorentz invariant value of 1500 TeV for the highest energy
synchrotron charges. By itself this is not impossible since, as
explained in Appendix~\ref{apsec:synch}, if $\eta$ is positive a
charge with lower energy can produce high frequency synchrotron
radiation. However, the Crab spectrum is well accounted for with a
single population of charges responsible for both the synchrotron
radiation and the IC $\g$-rays.  If there were enough charges to
produce the observed synchrotron flux with thirty times less energy
per electron, then the charges that do satisfy the \v{C}erenkov
constraint would presumably be at least equally numerous (since
their $\eta$ is smaller so helicity decay and source effects would,
if anything, produce more of them), and so would produce too many IC
$\g$-rays~\cite{JLMS}. Thus at least one of $\pm\eta_\pm$ must
satisfy {\it together} the synchrotron, synchrotron--\v{C}erenkov,
and the IC \v{C}erenkov constraints. That is, one of the four pairs
$(\pm\eta_\pm,\xi)$ must fall within the dashed bell-shaped boundary
in Fig.~\ref{fig:1}. This amounts to the statement that one of the
pairs $(\eta_\pm,\xi)$ must fall within the union of the bell-shaped
boundary and its reflection about the $\xi$ axis. This imposes the
two-sided upper bound $|\eta_\pm|<10^{-2}$ on one of the two
parameters $\eta_\pm$.

\subsection{Helicity decay}

The constraint $\Delta\eta=|\eta_+-\eta_-|<4$ on the difference
between the LV parameters for the two electron helicities was
deduced by Myers and Pospelov~\cite{MP} using a previous
spin-polarized torsion pendulum
experiment~\cite{Heckel}.\footnote{They also determined a
numerically stronger constraint using nuclear spins, however this
involves four different LV parameters, one for the photon, one for
the up-down quark doublet, and one each for the right handed up and
down quark singlets. It also requires a model of nuclear structure.}
Using  the photon decay bound $|\eta_\pm|<0.2$ discussed here, we
infer the stronger bound $\Delta\eta<0.4$. In this section we
discuss the possibility of improving on this constraint using the
process of helicity decay.

If $\eta_- > \eta_+$, negative helicity electrons are unstable to
decay into positive helicity electrons via photon emission. (We
assume for this section that $\eta_- > \eta_+$, the opposite case
works similarly. ) This reaction has no kinematic threshold.
However, the rate is very small at energies below an effective
threshold $(m^{2}M/\Delta\eta)^{1/3} \approx 10$ TeV (see Appendix
\ref{subsec:HelDec} for explicit rate calculations). The decay
lifetime is minimized at the effective threshold. Below that it is
{\it longer} than at least $\sim (\Delta\eta)^{-3}(p/10\, {\rm
TeV})^{-8}\times 10^{-9}$ seconds, while above it is given by $\sim
(p/10\, {\rm TeV})\times 10^{-9}$ seconds, independent of
$\Delta\eta$.

While accelerator energies are well below the effective threshold if
$\Delta\eta$ is $O(1)$, in principle one might still get a bound by
looking at fractional loss for a large population of polarized
accelerator electrons. In practice this is probably impossible due
to other polarizing and depolarizing effects in storage rings. But
just to see what would be required to improve the current bound
suppose (optimistically) that flipping of one percent of the
electrons stored for $10^4$ seconds could in be detected. The
experimental exclusion of this phenomenon would require that the
lifetime be greater than $10^6$ seconds. Using our overestimate of
the decay rate (\ref{eq:lowrate}) this would yield the constraint
$\Delta\eta<10^{3} (p/10\, {\rm GeV})^{-8/3}$. To improve on the
photon decay constraint would then require electron energies at
least around 200 GeV. When it was running LEPII produced 100 GeV
beams, but the currently highest energy electrons in storage rings
are around 30 GeV at HERA and 10 GeV at BABAR and the KEK B factory.
Hence even if the non-LV polarization effects could be somehow
factored out,  a useful helicity decay bound from accelerators does
not appear to be currently attainable.

A helicity decay bound can also be inferred using the charged
leptons in the Crab nebula. We have just argued that at least one of
the two parameters $\eta_\pm$ must lie within the dashed,
bell-shaped region together with its reflection about the $\xi$
axis.  Let us assume that the pair $(\eta_-,\xi)$ is inside the
allowed region. We can divide the region into three parts, A where
$\eta_->7 \cdot 10^{-8}$, B where $|\eta_-| \leq 7 \cdot 10^{-8}$,
and C where $\eta_-<-7 \cdot 10^{-8}$.  In the first case negative
helicity electrons can produce the synchrotron radiation, but
positive helicity positrons (which have LV parameter $-\eta_-$) are
below the synchrotron bound, so cannot. Now, if
 $\eta_+<-0.01$
neither positive helicity electrons or negative helicity positrons
can be responsible for the observed synchrotron radiation, since the
electrons cannot emit the necessary frequencies and the positrons
lose energy too rapidly via vacuum Cerenkov emission. Thus, in this
case, of the four populations of leptons only the negative helicity
electrons can produce the synchrotron radiation. However, an
$\eta_+$ this low would allow rapid helicity decay (since the
particle energy must be above 50 TeV which is around the effective
helicity decay threshold when $\Delta\eta=0.01$) from negative to
positive helicity electrons, leaving no charges to produce the
synchrotron. So if $\eta_-$ is in region A (which violates Lorentz
symmetry) we infer the lower bound $\eta_+>-0.01$. Similarly, if
$\eta_-$ is in region C, the same logic implies the upper bound
$\eta_+<0.01$. If $\eta_-$ is in region B, then negative helicity
electrons and positive helicity positrons are both able to produce
the synchrotron radiation. Whatever the value of $\eta_+$, at least
one of $\pm(\eta_--\eta_+)$ is positive, hence at least one of these
two species is always stable to helicity decay, so no helicity bound
can be presently inferred with one parameter in region B. If the
Crab spectrum could be modelled and observed precisely enough to
know that {\it both} (or all four) species must contribute to the
synchrotron radiation, a helicity bound could be obtained in case B.

\section{Other types of constraints on LV}
\label{other}

In this section, we briefly summarize and point to references for
various constraints on LV effects besides those associated with
$O(E/M)$ effects in QED.

\subsection{Constraints on dimension 3 and 4 operators}

For the $n=2$ term in (\ref{disprel1},\ref{disprel2}), the absence
of a strong threshold effect yields a constraint $\eta_2 \lesssim
(m/p)^2(M/\mu)$.  If we consider protons and put $\mu=m=m_p\sim 1$
GeV, this gives an order unity constraint when $p\sim \sqrt{mM} \sim
10^{19}$ eV. Thus the GZK threshold (see the following subsection),
if confirmed, can give an order unity constraint, but multi-TeV
astrophysics yields much weaker constraints.  The strongest
laboratory constraints on dimension three and four operators for
fermions come from clock comparison experiments using noble gas
masers~\cite{Bear:2000cd}. The constraints limit a combination of
the coefficients for dimension three and four operators for the
neutron to be below $10^{-31}$ GeV (the dimension four coefficients
are weighted by the neutron mass, yielding a constraint in units of
energy). This corresponds to a bound on $\eta_1$ of order $10^{-12}$
in the parametrization of (\ref{disprel2}) with $\mu=1$ GeV. For
more on such constraints see e.g.~\cite{AKbook,Bluhm:2003ne}.
Astrophysical limits on photon vacuum birefringence give a bound on
the coefficients of dimension four operators of
$10^{-32}$~\cite{KM}.

\subsection{Constraints at $O(E/M)$ from UHE cosmic rays}

In collisions of ultra high energy (UHE) protons with cosmic
microwave background (CMB) photons there can be sufficient energy in
the center of mass frame to create a pion, leading to the reaction
$p+\g_{\rm CMB}\rightarrow p+\pi^0$.
%
The so called GZK threshold~\cite{GZK} for the proton energy in this
process is
\beq E_{GZK}\simeq \frac{m_p m_\pi}{2E_\g} \simeq 3\times 10^{20}
{\rm eV}\times\left(\frac{2.7{\rm K}}{E_\g}\right) \label{EGZK} \eeq
To get a definite number we have put $E_\g$ equal to the energy of a
photon at the CMB temperature, 2.7{\rm K}, but of course the CMB
contains photons of higher energy. This process degrades the initial
proton energy with an attenuation length of about 50 Mpc. Since
plausible astrophysical sources for UHE particles are located at
distances larger than 50 Mpc, one expects a cutoff in the cosmic ray
proton energy spectrum, which occurs at around $5\times10^{19}$ eV,
depending on the distribution of sources~\cite{Stecker:2003wm}.

One of the experiments measuring the UHE cosmic ray spectrum, the
AGASA experiment, has not seen the cutoff. An
analysis~\cite{DeMarco:2003ig} from January 2003 concluded that the
cutoff was absent at the 2.5 sigma level, while another experiment,
HiRes, is consistent with the cutoff but at a lower confidence
level. (For a brief review of the data see~\cite{Stecker:2003wm}.)
The question should be answered in the near future by the AUGER
observatory, a combined array of 1600 water \v{C}erenkov detectors
and 24 telescopic air fluorescence detectors under construction on
the Argentine pampas~\cite{Auger}. The new observatory will see an
event rate one hundred times higher, with better systematics.

Many ideas have been put forward to explain the possible absence of
the GZK cutoff~\cite{Stecker:2003wm},  one being Lorentz violation.
According to equation~(\ref{EGZK}) the Lorentz invariant threshold
is proportional to the proton mass. Thus any LV term added to the
proton dispersion relation $E^2={\bf p}^2 + m_p^2$ will modify the
threshold if it is comparable to or greater than $m_p^2$ at around
the energy $E_{GZK}$. Modifying the proton and pion dispersion
relations, the threshold can be lowered, raised, or removed
entirely, or even an upper threshold where the reaction cuts off
could be introduced (see e.g.~\cite{Jacobson:2002hd} and references
therein).

If ultra-high energy cosmic rays (UHECR) are (as commonly assumed)
protons, then strong constraints on $n=3$ type dispersion can be
deduced from a) the absence of a vacuum \v{C}erenkov effect at GZK
energies and b) the position of the GZK cutoff if it will be
actually found.

a) For a soft emitted photon with a long wavelength, the partonic
structure of a UHECR proton is presumably irrelevant. In this case
we can treat the proton as a point particle as in the QED analysis.
With a GZK proton of energy $5 \times 10^{19}$ eV the constraint
from the absence of a vacuum \v{C}erenkov effect is $\eta<
O(10^{-14})$~\cite{Jacobson:2002hd}. Since the helicity of cosmic
rays is not observed, one can say only that this constraint must be
satisfied for at least one helicity.

For a hard emitted photon, the partonic nature of the proton is
important and the relevant mass scale will involve the quark mass.
The exact calculation considering the partonic structure for $n=3$
has not been performed. Treating the proton as structureless, the
threshold region would be similar to that in~\cite{Jacobson:2002hd}.
The allowed region in the $\eta-\xi$ plane would be bounded on the
right by the $\xi$ axis (within a few orders of magnitude of
$10^{-14}$) and below by the line $\xi=\eta$~\cite{Jacobson:2002hd}.
This constraint applies to both photon helicities, but only to one
proton helicity, since the UHECR could consist all of a single
helicity. In principle, however, what one can really constrain is
some combinations of the various quark dispersion parameters. This
approach has been worked out in detail using parton distribution
functions in~\cite{Gagnon:2004xh}.

b) If the GZK cutoff is observed in its predicted place, this will
place limits on the proton and pion parameters $\eta_p$ and
$\eta_\pi$. For example, if the GZK cutoff is eventually observed to
be somewhere between 2 and 7 times $10^{19}$ eV then there are
strong constraints of $O(10^{-11})$ on the relevant $\eta_p$ and
$\eta_\pi$~\cite{Jacobson:2002hd}. (Allowing for helicity
dependence, no set of parameters allowing long distance propagation
(forbidding the vacuum \v{C}erenkov effect) should modify the GZK
cutoff.)

As a final comment, an interesting possible consequence of LV is
that with upper thresholds, one could possibly reconcile the AGASA
and Hi-Res/Fly's Eye experiments.  Namely, one can place an upper
threshold below $10^{21}$ eV while keeping the GZK threshold near
$5\times 10^{19}$ eV.  Then the cutoff would be ``seen'' at lower
energies but extra flux would still be present at energies above
$10^{20}$ eV, potentially explaining the AGASA
results~\cite{Jacobson:2002hd}.  The region of parameter space for
this scenario is terribly small, however, again of $O(10^{-11})$.

\subsection{Constraints on dimension 6 operators}

As previously mentioned, CPT symmetry alone could exclude the
dimension five LV operators in QED that give $O(E/M)$ modifications
to particle dispersion relations. Moreover, the constraints on those
have become quite strong. Hence we close with a brief discussion of
the constraints that might be possible at $O(E^2/M^2)$. Such LV
effects arise for example from dimension 6 operators. Note that
helicity dependence of the LV parameters is not required in this
case, and on the other hand it can occur without violating CPT.
Without any particular theoretical prejudice, one should keep in
mind that constraints will generally only limit the parameters for
{\it some} helicity species, while they might be evaded for other
helicities.

As previously discussed we can estimate that the LV modification of
the dispersion relation becomes important when comparable with the
mass term, $\eta_4 p^4/M^2 \lesssim m^2$, which yields
\beq \eta_4\lesssim \left(\sqrt{\frac{m}{1\, {\rm
MeV}}}\frac{10^{17}\, {\rm eV}}{p}\right)^4. \label{eta4bound} \eeq
Thus, for electrons, an energy around $10^{17}$ eV is needed for an
order unity constraint on $\eta_4$, and we are probably not going to
see any effects directly from such electrons.

For protons an energy $\sim 10^{18}$ eV is needed. This is well
below the UHE cosmic ray energy cutoff, hence if and when
Auger~\cite{Auger} confirms the identity of UHE cosmic rays as
protons at the GZK cutoff, an impressive constraint of order
$\eta_4\lesssim 10^{-5}$ will follow from the absence of vacuum
\v{C}erenkov radiation for $10^{20}$ eV protons. From the fact that
the GZK threshold is not shifted, a constraint of order
$\eta_4\gtrsim -10^{-2}$ will follow, assuming equal $\eta_4$ values
for proton and pion.

In fact, if one assumes the cosmic rays already observed near but
below the GZK cutoff are hadrons, one obtains a strong
bound~\cite{Gagnon:2004xh}. Depending on the species and helicity
dependence of the LV coefficients, bounds of order $10^{-2}$ or
better can be placed on $\eta_4$.  The bounds claimed in
~\cite{Gagnon:2004xh} are actually two sided, and come about in a
manner somewhat analogous to (but more complicated than) the photon
decay constraint (see section~\ref{photondecay}). They are derived
by using a parton model for particles where the LV coefficients
apply to the constituent partons.  By considering many different
outgoing particle spectra from an incoming hadron in combination
with the parton approach the authors of~\cite{Gagnon:2004xh} are
able to find sets of reactions that yield two sided bounds. Hence,
the parton approach is quite useful, as it dramatically increases
the number of constraints that can be derived from a single incoming
particle.

One might think that impressive constraints can also be obtained
from the absence of neutrino vacuum \v{C}erenkov radiation: putting
in 1 eV for the mass in (\ref{eta4bound}) yields an order unity
constraint from 100 TeV neutrinos, but only if the \v{C}erenkov {\it
rate} is high enough. The rate will be low, since it proceeds only
via the non-local charge structure of the neutrino. Recent
calculations~\cite{Dave?} have shown that the rate is not high
enough at that energy, even for cosmogenic neutrinos. However, for
$10^{20}$ eV UHE neutrinos, which may be observed by the proposed
EUSO (Extreme Universe Space Observatory)~\cite{EUSO} and/or OWL
(Orbiting Wide-angle Light collectors)~\cite{OWL} satellite
observatories, the rate will be high enough to derive a strong
constraint as long as the neutrinos are cosmogenic, and perhaps even
if they originate closer to the earth. The specific value of the
constraint would depend on the exact value of the rate, which has
not yet been computed. For a {\it gravitational} \v{C}erenkov
reaction, the rate (which is lower but easier to compute than the
electromagnetic rate) would be high enough for a $10^{20}$ eV
neutrino from a distant source to radiate provided $\eta_4\gtrsim
10^{-2}$. Hence in this case one might obtain a constraint of order
$\eta_4\lesssim 10^{-2}$ from gravitational \v{C}erenkov. If
\v{C}erenkov constraints apply only when the observed UHE neutrino
originates from a \textit{distant} source, one would need to either
identify an astrophysical object as the source or somehow otherwise
rule out local generation of the neutrino.

A time of flight constraint at order $(E/M)^2$ might be
possible~\cite{GACsecgen} if gamma ray bursts produce UHE ($\sim
10^{19}$ eV) neutrinos, as some models predict, via limits on time
of arrival differences of such UHE neutrinos vs. soft photons (or
gravitational waves). Another possibility is to obtain a vacuum
birefringence constraint with higher energy photons~\cite{Mitro},
although such a constraint would be less powerful since EFT does not
imply that the parameters for opposite polarizations are opposite at
order $(E/M)^2$. If future GRB's are found to be polarized at $\sim
100$ MeV, that could provide a birefringence constraint
$|\xi_{4+}-\xi_{4-}|\lesssim 1$.

\section{Future prospects}
\label{conclusions}

In the last decade or so the old dogma that all our observations are
insensitive to Planck scale effects has been shown to be quite wrong
if Lorentz symmetry is violated. A large number of impressively
strong constraints have been obtained on LV Planck scale effects
arising from operators of mass dimension three, four, five, and even
six.

{}From the conceptual point of view, the most burning issue is
naturalness of small low energy LV. We have explained in this
article both the rationale for adopting an effective field theory
parametrization of LV, and the ``naturalness" problem of preserving
approximate low energy Lorentz invariance if LV is to exist in the
UV theory.  As discussed in section~\ref{naturalness}, currently the
best prospect for resolving this issue in favor of LV is via
supersymmetry, but perhaps other ideas could work.  In the Lorentz
violating, softly broken SUSY QED framework of
Ref.~\cite{Bolokhov:2005cj}, dimension five (CPT violating) LV
operators are already strongly constrained by current QED
experiments, but the dimension six LV operators are not yet
constrained at the $M_{Planck}^{-2}$ level. However the authors of
that reference indicate that Planck level constraints on dimension
six LV quark operators in the standard model may be possible. We
emphasize that high energy LV dispersion of the sort discussed in
the present paper does not occur in the scenario of
Ref.~\cite{Bolokhov:2005cj}.

Constraints continue to improve with advances in observational
technology. In this concluding section we examine some of the future
prospects for improving the constraints on high energy LV using
astrophysical observations. To begin, we note that while photon
``time of flight" constraints are not competitive with other current
constraints, they enjoy a special status in being less dependent on
assumptions about the underlying theoretical framework.  Given that
time of flight constraints are strongest with a high energy signal
that has structure on short time scales, gamma ray bursts are
probably the best objects for improving these constraints.
Experiments like GLAST (Gamma-ray Large Area Space
Telescope)~\cite{Glast-web} may be helpful in improving bounds from
time of flight; they might lead to a better understanding of GRB
emissions and provide far better photon counts and time resolution
(less than $10\mu$sec). These improvements might help overcome some
of the problems involving GRB time of flight measurements discussed
in section~\ref{sec:tof}.  See e.g.~\cite{Piran:2004qe} for a
detailed discussion of these issues.

Here we focused on LV in the dimension five, CPT violating sector of
QED, characterized by the photon parameter $|\xi|$ bounded at
$O(10^{-4})$ and the electron parameters $|\eta_\pm|$ bounded
collectively at $O(10^{-1})$, while at least one of them is bounded
at $O(10^{-2})$. The most promising prospect for significant
improvement in these bounds would be a closer study of the effect of
this sort of LV on the complete electromagnetic spectrum of the Crab
nebula and or other plerions. We have been very conservative in
imposing constraints, by allowing for the possibility that with LV
the synchrotron radiation could be produced by much lower energy
charges than required in the usual Lorentz invariant case. We have
also tolerated the possibility of four different populations of
charges behaving differently (two helicities for electrons and
positrons). By modelling more completely the production of the
electron-positron wind and electromagnetic spectrum in the presence
of such LV it may be possible to infer that charges of energy 1500
TeV are in fact required as in the Lorentz invariant case. If so,
the constraints on $\eta_\pm$ might be improved by a factor of order
$\sim (50/1500)^3\sim 10^{-4}$ to $|\eta_\pm|\lesssim 10^{-6}$, and
that on $|\xi|$ by a factor of order $\sim 10^{-2}$ to
$|\xi|\lesssim 10^{-6}$ as well. This would seem to be rather
definitive so is a useful goal. To achieve it would perhaps require
not only better modelling of the LV effects but more precise
observations of the Crab nebula. In this regard we can look forward
to the observations to be made with GLAST which should start its
observations in 2007.

The birefringence bound, currently of order $|\xi| \lesssim 10^{-4}$
from UV polarization observations of a distant galaxy, might be
improved in various ways. Since the LV effect scales as the square
of the energy, the most dramatic improvement would be if polarized
gamma rays from a distant source could be observed, for example from
a GRB. (The observation reported in~\cite{CB03} appears now to be
unreliable.) This might be possible with RHESSI or via the IBIS
camera of the INTEGRAL satellite~\cite{Integral}, using the
so-called Compton mode that was used in the previous RHESSI
analyses. Less dramatically, it might instead be possible to improve
by an order of magnitude the constraint reported in~\cite{GK},
obtained using UV light from the radio galaxy 3C 256, via
specifically targeted observations possibly using FORS1~\cite{Fors1}
at the ESO Very Large Telescope (VLT), and/or by observing more
distant sources of polarized UV light.

Bounds on LV effects on threshold reactions for ultra-high energy
cosmic rays can be significant even when suppressed by two powers of
the Planck mass, i.e. at order $E^2/M^2$. This is important since
CPT symmetry alone  could preclude effects at order $E/M$. We
already discussed the currently available bounds. The possibility of
improvement lies in experiments which will be capable of reaching
high sensitivities at the small fluxes associated with the highest
energy particles. The Auger detector will be the next big thing,
with first release of data expected in summer 2005.

It is conceivable that ultra-high energy neutrinos could provide
extremely stringent constraints on LV due to their small mass. The
vacuum \v{C}erenkov effect occurs at a given energy with a minimum
LV coefficient that scales with mass as $m^2$. On the other hand the
rate for the reaction is very low, so it is not clear whether in
practice this possibility will pan out. UHE neutrinos might be
observed using IceCube~\cite{IceCube}, or planned detectors such as
the cited satellites EUSO~\cite{EUSO} and OWL~\cite{OWL} (the former
being already in an advanced stage of development). Also promising
are upcoming experiments like the sea based detector NESTOR
(Neutrino Extended Submarine Telescope with Oceanographic
Research)~\cite{NESTOR} which is now starting its activity, or the
balloon-borne detector ANITA (Antarctic Impulse Transient
Antenna)~\cite{ANITA} whose first flight is planned for the end of
2006. Finally also very effective could be some proposed experiments
like the underground salt based detector SalSA (Salt-dome Shower
Array)~\cite{SALSA}.

In closing, we have been motivated by considerations of quantum
gravity to look for evidence of, and bounds on, Lorentz violation.
These considerations include suggestions of LV from various
viewpoints, as well as simple desperation for observational
guidance. The constraints discussed here and elsewhere are mounting
up to a significant limitation on the possibility of LV in quantum
gravity, at least insofar as would appear in low energy effective
field theory. It seems very worthwhile to push on, even if the
ultimate result will only be extremely restrictive constraints. In
this way we will have solidified the observational foundations of
Lorentz symmetry, and acquired valuable constraints on future
speculations on the nature of quantum gravity.

\appendix

\section{LV cutoff of synchrotron radiation frequency}
\label{apsec:synch}

Cycling electrons in a magnetic field $B$ emit synchrotron radiation
with a spectrum that sharply cuts off at a frequency $\omega_c$
given in the Lorentz invariant case by the formula
\begin{equation}
\omega_c=\frac{3}{2} eB\frac{\gamma^3(E)}{E}\, , \label{eq:opeaklv}
\end{equation}
where $\gamma(E)=(1-v^2(E)/c^2)^{-1/2}$. Here $v(E)$ is the electron
group velocity, and $c$ is the usual low energy speed of light. In
standard relativistic physics, $E=\g m$, so the energy dependence in
(\ref{eq:opeaklv}) is entirely through the factor $\g^2$, which
grows without bound as the energy grows. Arbitrarily large
synchrotron frequencies are therefore possible.

In the Lorentz violating case with negative values of $\eta$,
electrons have a maximal group velocity strictly less than the low
energy speed of light, hence there is a maximal synchrotron
frequency that can be produced, regardless of the electron
energy~\cite{Crab}.\footnote{Since we have LV in the photon sector
as well, there is no ``speed of light'' per se.  However, the
emitted frequencies of synchrotron radiation are much lower than the
energy of the source particles in the Lorentz invariant case.
Effectively the LV violation in the photon sector can be ignored.}
As we shall argue below, Eq. (\ref{eq:opeaklv}) still  holds in this
case, assuming the framework of effective field theory. Therefore
the maximal frequency is obtained by maximizing $\gamma^3(E)/E$
with respect to the electron energy. Using the difference of group
velocities
\begin{equation}
c-v(E)\simeq \frac{m^2}{2E^2} - \eta\, \frac{E}{M}, \label{eq:vdiff}
\end{equation}
obtained from the electron dispersion relation
(\ref{eq:dispfermionhighE}), we find that this maximization yields
\begin{equation}
\omega_c^{\rm max}=0.34 \, \frac{eB}{m}(-\eta m/M)^{-2/3}.
\label{eq:opeaklv2}
\end{equation}
This maximum frequency is attained at the energy $E_{\rm
max}=(-2m^2M/5\eta)^{1/3}=10\, (-\eta)^{-1/3}$ TeV. This is higher
than the energy that produces the same cutoff frequency in the
Lorentz invariant case, but only by a factor of order unity.

Note that if $\eta$ is {\it positive}, then the effect is the
opposite: an electron can produce a given frequency of synchrotron
radiation with an energy {\it less} than in the Lorentz invariant
case. The electron speed can even exceed the low frequency speed of
light, at which point $\gamma(E)$ diverges. This corresponds to the
soft \v{C}erenkov threshold discussed in
section~\ref{sebsec:cheren-thr}.

We shall now justify the derivation of (\ref{eq:opeaklv}) in the LV
case, adding some details to previously published work\cite{Crab}.
The first step is a {\it purely kinematical} analysis that does not
assume Lorentz invariance, which follows the standard heuristic
derivation~\cite{Jackson}. A cycling electron of energy $E$ emits
radiation in a cone of some opening angle $\delta(E)$. The cone
sweeps past a distant observer as the electron moves on a circle of
radius $R(E)$
 through an angle $\delta(E)$. The electron travels
at a speed (group velocity) $v(E)$  so the time it takes to orbit
through the angle $\delta(E)$ is $\Delta t=R(E)\delta(E)/v(E)$. The
light from the leading edge of the cone travels a distance
$c(\omega)\Delta t$ while the electron travels the distance
$v(E)\Delta t$ toward the observer. Hence the spatial width of the
pulse seen by the observer is approximately $(c(\omega)-v(E))\Delta
t$, which arrives at the observer over a time interval equal to this
distance divided by the speed of light. The cut off frequency of the
synchrotron pulse is roughly the inverse of this time interval,
\begin{equation}
\omega_c=\frac{3}{4} \frac{1}{R(E)\delta(E)}\;
\frac{1}{c(\omega_c)-v(E)},\label{eq:opeaktilde}
\end{equation}
where we have used the fact that the electron and photon speeds are
very close to the low energy speed of light $c$, which is set equal
to unity. In the Lorentz invariant case the radius is given by
$R(E)=E/eB$ and the opening angle is $\delta(E)\sim\gamma^{-1}(E)$.
The numerical constant in (\ref{eq:opeaktilde}) is chosen so that
when these values are substituted the correct relativistic result
(\ref{eq:opeaklv}) is obtained.

Under the assumption that the Lorentz violation is described by
effective field theory, we will argue that $R(E)$ and $\delta(E)$
are the same in the LV case as in the Lorentz invariant case.
Moreover, since the emitted photons have relatively low energy
compared to the electrons, it turns out that $\xi$ can be neglected
in the relevant region of parameter space. (As shown in~\cite{Crab}
the photon energy is low enough to neglect any possible LV
correction as long as $|\xi|\lesssim 10^{11}(-\eta)^{4/3}$.)  Thus
$c(\omega_c)$ in (\ref{eq:opeaktilde}) can be replaced by $c$, so
the reciprocal of the difference of group velocities is well
approximated by $2\gamma^2(E)$, as in the Lorentz invariant case.
This yields (\ref{eq:opeaklv}) where now all the dependence on the
LV is in how the gamma factor depends on $E$.

The radius $R(E)$ is determined by the equation of motion for the
electron in a magnetic field. All Lorentz violating terms in the
equation of motion are suppressed by the ratio $E/M$. From this it
is clear that the trajectory determined by a given initial position
and momentum is not much affected by the Lorentz violation, as long
as $E$ remains much smaller than the Planck energy. \footnote{A
subtlety here is that this argument fails if we refer
 to the velocity rather than to the momentum.
The reason is that tiny differences in the speed make a huge
difference in the momentum and energy when the speed is close to the
speed of light.}

This conclusion about $R(E)$ can be more explicitly verified by
reference to the modified Hamilton's equations, with the Hamiltonian
given in terms of the momentum by the dispersion relation between
energy and momentum. The leading high energy corrections come from
modifications to the minimal coupling terms. As usual the minimal
coupling is incorporated replacing the momentum by ${\bf p} - e{\bf
A}$, where ${\bf A}$ is a vector potential for the magnetic field.
This yields the equation of motion ${\bf a}=[1 + 3\eta E/2M](e/E)\,
{\bf
  v}\times{\bf B}$, where we have kept only the lowest
order term in $\eta$ and assumed relativistic energy $E\gg m$. Since
$E\ll M$, the presence of the Lorentz violation makes very little
difference to the orbital equation, hence we conclude that to a very
good approximation the radius is related to the magnetic field and
the energy of the electron by the standard formula $R(E)= E/eB$
(where again the speed of the electron has been set equal to unity).
This result was independently confirmed by~\cite{Ellis:2003sd} from
the starting point of the Dirac equation.

The angle $\delta(E)$ scales in the Lorentz invariant case as
$\gamma^{-1}(E)$. Since $R(E)$ and hence the charge current is
nearly unaffected by the Lorentz violation, any significant
deviation in $\delta(E)$ could only come from the modified response
of the electromagnetic field to the given current. The Lagrangian in
the presence of the dimension five LV operator for the EM field in
(\ref{dim5}) takes the form
\be L=\frac{1}{4}F_{ab}F^{ab}+\frac{\xi}{M}G(A)+A_aj^a, \ee
where $\xi G(A)/M$ is the LV operator and $j^a$ is the given
current. The equations of motion are
\be
\partial_aF^{ab}-\frac{\xi}{M}\frac{\partial G(A)}{\partial A_b}=J^b.
\ee
If we expand $A^a=A^a_0+A^a_1$, where $A^a_0$ is the field in the
Lorentz invariant case, then to lowest order the LV correction
$A^a_1$  satisfies
\be
\partial_aF_1^{ab}=\frac{\xi}{M}\frac{\partial G(A^c_0)}{\partial A_a}.
\label{pert} \ee
The correction $A^a_1$ is suppressed by at least one power of $M$.
In terms of Fourier components, this suggests that the relative size
of the LV correction is of order $\xi \omega/M$, which is extremely
small for even the peak photon energies $\approx 10^8$ eV present in
the synchrotron radiation from the Crab nebula. Therefore near the
source current the amplitude of the electromagnetic field
modification is negligible in comparison to the zeroth order
synchrotron radiation (which is still present), and hence
$\delta(E)$ scales with $\gamma^{-1}(E)$ in the usual way. Note that
this does not mean that the relative size of the correction is small
everywhere. The source of $A^a_1$ on the right hand side of
(\ref{pert}) is non-vanishing not just where the current $j^b$ lies
but also wherever the emitted radiation propagates. It can therefore
produce a cumulative effect that can be large. In fact, this must
clearly happen, since the group velocity is modified by the LV term,
so that after enough propagation the relative correction must become
of order unity so that it can shift the support of the radiation to
a disjoint location.

A detailed analysis of synchrotron radiation in this LV QED theory
was carried out in Ref.~\cite{Montemayor:2004mh}. Our heuristically
obtained results were confirmed there, and certain regimes were
identified in which the LV solution deviates strongly from the
Lorentz invariant one. Our interpretation of such deviations is what
was mentioned at the end of the previous paragraph, i.e. that these
are cumulative effects. A claim of significant deviations was made
in Ref.~\cite{Castorina:2004rc} however we do not currently
understand the basis of that claim.

\section{Threshold configurations}
\label{apsec:threshconfig}

Threshold configurations and new phenomena in the presence of LV
dispersion relations were systematically investigated
in~\cite{CGlong,Mattingly:2002ba} (see also references therein). We
give here a brief summary of the results. We shall consider
reactions with two initial and two final particles (results for
reactions with only one incoming or outgoing particle can be
obtained as special cases).  Following our previous choice of EFT we
allow each particle to have an independent dispersion relation of
the form (\ref{gen-disprel}) with $E(p)$ a  monotonically increasing
non-negative function of the magnitude $p$ of the 3-momentum $\bf
p$.  While the assumption of monotonicity could perhaps be violated
at the Planck scale, it is satisfied for any reasonable low energy
expansion of a LV dispersion relation. EFT further implies that
energy and momentum are additive for multiple particles, and
conserved.

Consider a four-particle interaction where a target particle of
3-momentum ${\bf p}_2$ is hit by a particle of 3-momentum ${\bf
p}_1$, with an angle $\alpha$ between the two momenta, producing two
particles of momenta ${\bf p}_3$ and ${\bf p}_4$. We call $\beta$
the angle between ${\bf p}_3$ and the total incoming 3-momentum
${\bf p}_{in}={\bf p}_1+{\bf p}_2$. We define the notion of a
threshold relative to a fixed value of the magnitude of the target
momentum $p_2$. A lower or upper threshold corresponds to a value of
$p_1$ (or equivalently the energy $E_1$) above which the reaction
starts or stops being allowed by energy and momentum conservation.

We now introduce a graphical interpretation of the energy-momentum
conservation equation that allows the properties of thresholds to be
easily understood. For given values of $(p_1, p_2,\alpha,\beta,
p_3)$, momentum conservation determines $p_4$. Since $p_3$ and $p_4$
determine the final energies $E_3$ and $E_4$, we can thus define the
final energy function $E_f^{\alpha,\beta,p_3}(p_1)$. (Since $p_2$ is
fixed we drop it from the labelling.) Energy conservation requires
that $E_f$ be equal to $E_i(p_1)$, the initial energy (again, we do
not indicate the dependence on the fixed momentum $p_2$).

\begin{figure}[htb]
\vbox{ \hfil \scalebox{0.60}{{\includegraphics{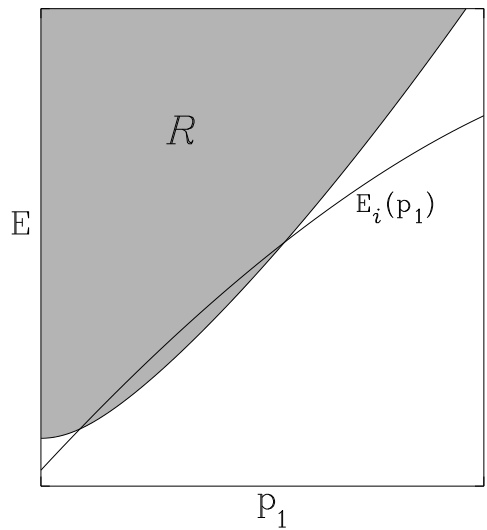}}} \hfil }
\bigskip
\caption{Graphical representation of energy-momentum conservation in
a two-particle reaction. $\mathcal{R}$  is the region covered by all
final energy curves $E_f^{\alpha,\beta,p_3} (p_1)$ for some fixed
$p_2$, assuming momentum conservation holds to determine $p_4$. The
curve $E_i(p_1)$ is the initial energy for the same fixed $p_2$.
Where the latter curve lies inside  $\mathcal{R}$ there is a
solution to the energy and momentum conservation equations. In the
example shown there is both a lower and an upper threshold for the
reaction.} \label{fig:R}
\end{figure}

Now consider the region $\mathcal{R}$ in the $(E, p_1)$ plane
covered by plotting $E_f^{\alpha,\beta,p_3}(p_1)$  for all possible
configurations  $(\alpha,\beta, p_3)$. An example is shown in
Figure~\ref{fig:R}. The region $\mathcal{R}$ is bounded below by
$E=0$ since the particle energies are assumed non-negative, hence it
has some bounding curve $E_B(p_1)$. Similarly one can plot
$E_i(p_1)$. The reaction threshold occurs when $E_i(p_1)$ enters or
leaves $\mathcal{R}$, since it is precisely in $\mathcal{R}$ that
there is a solution to the energy and momentum conservation
equations.

This graphical representation demonstrates that in any threshold
configuration (lower or upper) occurring at some $p_1$, the
parameters $(\alpha,\beta, p_3)$ are such that the final energy
function $E_f^{\alpha,\beta,p_3}(p_1)$ is minimized. That is, the
configuration always yields the minimum final particle energy
configuration conserving momentum at fixed $p_1$ and $p_2$. From
this fact, it is easy to deduce two general properties of these
configurations:
\begin{enumerate}
\item All thresholds for processes with two outgoing particles
occur at parallel final momenta ($\beta=0$). \item For a two
particle initial state the momenta are antiparallel at threshold
($\alpha=\pi$).
\end{enumerate}
These properties are in agreement with the well known case of
Lorentz invariant kinematics. Nevertheless, LV thresholds can
exhibit new features not present in the Lorentz invariant theory, in
particular upper thresholds, and asymmetric pair creation.

Figure~\ref{fig:R} clearly shows that LV allows for a reaction to
not only to start at some lower threshold but also to end at some
upper threshold where the curve $E_i$ exits the region
$\mathcal{R}$. It can even happen that $E_i$ enters and exits
$\mathcal{R}$ more than once, in which case there are what one might
call ``local" lower and upper thresholds.

Another interesting novelty is the possibility to have a (lower or
upper) threshold for pair creation with an unequal partition of the
initial momentum $p_{in}$ into the two outgoing particles (i.e.
$p_3\neq p_4\neq p_{in}/2$). Equal partition of momentum is a
familiar result of Lorentz invariant physics, which follows from the
fact that the final particles are all at rest in the zero-momentum
frame at threshold. This has often been (erroneously) presumed to
hold as well in the presence of LV dispersion relations.

\begin{figure}[htb]
\bigskip
\vbox{ \hfil \scalebox{0.60}{{\includegraphics{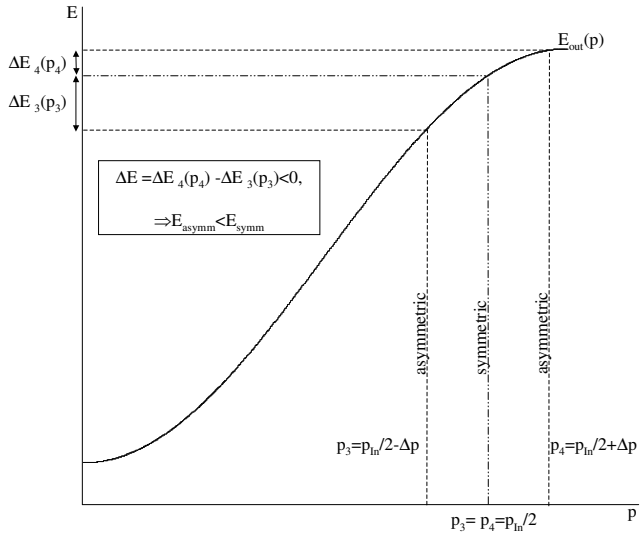}}} \hfil }
\bigskip
\caption{ Asymmetric pair production. The negative curvature of the
outgoing particle dispersion relation allows to save energy by
providing the pair partners with different portions of the initial
momentum $p_{in}$. } \label{fig:asymm}
\end{figure}

A reason for the occurrence of asymmetric LV thresholds can be seen
graphically, as shown in Figure~\ref{fig:asymm}. Suppose the
dispersion relation for a massive outgoing particle $E_{out}({\bf
p})$ has negative curvature at $p=p_{in}/2$, as might be the case
for negative LV coefficients. Then a small momentum-conserving
displacement from a symmetric configuration can lead to a net
decrease in the final state energy. According to the result
established above, the symmetric configuration cannot be the
threshold one in such a case. A lower $p_1$ could satisfy both
energy and momentum conservation with an asymmetric final
configuration.
A {\it sufficient} condition for the pair-creation threshold
configuration to be asymmetric is that the final particle dispersion
relation has negative curvature at $p=p_{in}/2$. This condition is
not necessary however, since it could happen that the energy is
locally but not globally minimized by the symmetric configuration.

\section{Rates and thresholds of LV processes}
\label{apsec:rates}

In this appendix we first derive a general expression for reaction
rates. We then discuss the rates and thresholds for three reactions
of interest: photon decay, vacuum \v{C}erenkov, and helicity decay.

\subsection{General rate expressions}

The first step in deriving constraints from particle interactions is
to derive the rate $\Gamma$ for various processes.  As it turns out,
the most useful constraints are derived from reactions where a
single particle decays into two particles.  If we label the incoming
particle momentum and helicity/polarization by ${\bf p},s$ and the
final particles by ${\bf p}',s',{\bf p}'', s''$ then the rate for
fixed helicities is given by~\cite{Peskin}
\begin{eqnarray} \label{eq:peskinrate}
\Gamma(p,s,s',s'') = \int \frac {1} {8 E_{p,s}} \frac {d^3{p}'
d^3{p}''} {(2 \pi)^2 E'_{p',s'} E''_{p'',s''}} |{\sf M}({\bf
p},s,{\bf p}',s',{\bf p}'',s'')|^2 \\ \nonumber \times
\delta(E_{p,s}-E'_{p',s'} - E''_{p'',s''}) \delta^{(3)}({\bf p}-{\bf
p}'-{\bf p}'')
\end{eqnarray}
where ${\sf M}({\bf p},s,{\bf p}',s',{\bf p}'',s'')$ is the matrix
element.  As mentioned in section \ref{subsubsec:fermions} our fermion
spinors are not quite normalized to $2E$.  As a result
(\ref{eq:peskinrate}) is not quite correct. However, as long as we
work in the preferred frame, this correction is small and can be
neglected since it is only a factor in overall normalization.

If we choose the initial particle to be travelling in the
$z$-direction, then we can integrate over ${\bf p}''$ and apply the
axial symmetry to get
\begin{equation}
\Gamma(p,s,s',s'') = \int \frac {1} {16 \pi} \frac {dp'_z p'_\bot
dp'_\bot} {E_{p,s} E'_{p',s'}  E''_{{\bf p}-{\bf p}',s''}} |{\sf
M}|^2 \delta(E_{p,s}-E'_{p',s'}-E''_{{\bf p}-{\bf p}',s''}).
\end{equation}
Integrating over $p'_\bot$ then yields
\begin{equation} \label{eq:gendedt1}
\Gamma(p,s,s',s'') =  \int_{p'_{z1}}^{p'_{z2}} \frac {1} {16 \pi}
\frac {p'_\bot dp'_z} {E_{p,s} E'_{p',s'} E''_{{\bf p}-{\bf
p}',s''}} |{\sf M}|^2 \bigg{|} \frac {\partial (E'_{p',s'} +
E''_{{\bf p}-{\bf p}',s''})} {\partial p'_\bot} \bigg{|}^{-1}.
\end{equation}
This expression and those that follow are evaluated at the solution
of the energy conservation equation,
$p'_\bot=p'_\bot(p'_z,p,s,s',s'')$. The limits of integration
$p'_{z1}, p'_{z2}$ are the bounds between which such a solution
exists.  We are mostly interested in cases where there is no
solution unless $p$ is greater than some threshold momentum
$p_{th}$.

For reactions that do not occur in the Lorentz-invariant limit, the
phase space must close up as we approach Lorentz invariance, i.e.\
the magnitude of $p'_\bot$ vanishes as $E_{Pl} \rightarrow \infty$.
Hence $p'_\bot$ must be small relative to $p'_z$ and $p-p'_z$ if
$p\ll E_{Pl}$. The exception to this is if either $p'_z,p-p'_z$ is
near zero, i.e. when either $p'_{z1} \approx 0$ or $p'_{z2} \approx
p$. This region is, however, a negligible amount of phase space and
so we ignore it. Thus $p'_\bot$ can be treated as an expansion
parameter.

Expanding the derivative of $E'_{p',s'}+E''_{{\bf p}-{\bf p}',s''}$
with respect to $p'_\bot$ gives
\begin{equation}
\bigg{|} \frac{\partial (E'_{p',s'} + E''_{{\bf p}-{\bf p}',s''})}
{\partial p'_\bot}\bigg{|}^{-1}\approx \frac {E'_{p',s'} E''_{{\bf
p}-{\bf p}',s''}} {p'_\bot E_{p,s}},
\end{equation}
and substituting into (\ref{eq:gendedt1}) gives
\begin{equation} \label{eq:gendedt}
\Gamma(p,s,s',s'') =  \int_{p'_{z1}}^{p'_{z2}} \frac {1} {16 \pi}
\frac {dp'_z} {E_{p,s}^2} |{\sf M}|^2.
\end{equation}
To compute an energy loss rate for the \v{C}erenkov effect a $p'_z$
must be inserted in the integrand to reflect the energy carried off
by the emitted photon.

\subsection{Photon decay matrix element} \label{subsubsec:photondecayrate}

We now show that, if the incoming momentum $p$ is above the
threshold $p_{th}$, then the rate for photon decay is very high.

The QED matrix element for photon decay is
\begin{equation} \label{eq:ME}
{\rm i}{\sf M}={\rm i}e \bar{u}_{s}(p') \epsilon^{\pm}_\alpha
\gamma^\alpha v_s(p'')
\end{equation}
where $\bar{u}(p')$ is the outgoing electron spinor and $v(p'')$ is
the outgoing positron spinor. We consider the case where the
electron and positron have positive and negative helicity
respectively, so only the coefficient $\eta_+$ is involved. The LV
term in the electron dispersion is $\eta_+ p^3$ (with $u^+$
wavefunction (\ref{spinors})) while that in the positron dispersion
is $-\eta_+ p^3$ (with $v^-$ wavefunction (\ref{pspinors})). The
result is directly transferable to the $\eta_-$ case. We will also
relabel $E(p)$ as $\omega(k)$ since the incoming particle is a
photon.

With these choices the dominant contribution to the matrix element
at high energy is straightforwardly evaluated as
\begin{equation}
{\rm i}{\sf M}={\rm i} e \sqrt{2 E'} \sqrt {2E''}
\left[\chi_+(p')^\dagger ( \vec{\epsilon}_{\pm} \cdot
{\vec{\sigma}}) \chi_+(p'')\right]
\end{equation}
which is equal to
\begin{equation} \label{eq:Mphotdecaytheta}
{\rm i}e \sqrt{2E'E''} \left[ \cos\left(\frac {\theta''} {2}\right)
\sin \left(\frac {\theta'} {2}\right) (1 \pm 1) -  \sin
\left(\frac{\theta''} {2}\right) \cos \left(\frac {\theta'}
{2}\right) (1 \mp 1) \right]
\end{equation}
where $\theta',\theta''$ are the opening angles of the electron and
positron and the $\pm$ reflects the initial photon polarization.
(The overall phase depends on the phase convention for the spinors
and does not affect the result.) Note that the matrix element
vanishes in the threshold configuration when the momenta are all
parallel. This is because we chose the electron and positron
helicities to be opposite. Nevertheless, above threshold the rate is
high enough to obtain a useful constraint, as we now argue.

The opening angles are small, since the perpendicular momenta 
are
Planck suppressed. Therefore we can expand
(\ref{eq:Mphotdecaytheta}) to first order in $\theta',\theta''$,
yielding
\beq \label{eq:phot-dec-M} {\rm i}{\sf M}={\rm
i}e\sqrt{E'E''/2}\left[\frac {p'_\bot} {p'_z} (1 \pm 1) - \frac
{p'_\bot} {p''_z} (1 \mp 1)\right], \eeq
where $p'_\bot$ is the magnitude of the 
perpendicular momentum (which is the same for 
the electron and positron).
Recall that $p'_\bot$ is determined from the energy conservation
equation (where $k$ is the initial photon momentum),
\begin{equation} \label{eq:cons}
\pm  \frac{\xi k^2}{M}=\frac {m^2} {p'_z} + \frac {(p'_\bot)^2} {p'_z} +
\frac{\eta_+ (p'_z)^2}{M} + \frac {m^2} {k-p'_z} + \frac {(p'_\bot)^2}
{k-p'_z}-\frac{\eta_+ (k-p'_z)^2}{M}.
\end{equation}
In (\ref{eq:cons}) we have
already cancelled the 0th order terms using momentum conservation,
and we have neglected higher order terms in $p'_\bot/p_z$.
We have also assumed that the $z$-components of the electron
and positron momenta are both positive. This is not necessarily
the case, but when $m\ll k\ll M$ (which has also been assumed)
we are only missing a tiny fraction of the phase space by
making this restriction.

Defining $z=p'_z/k\in [0,1]$, $p'_\bot$ is given by
\beq p^{'2}_\bot=[\pm \xi - \eta_+(2z-1)] z(1-z)\frac{k^3}{M} - m^2.
\label{eq:pbot} \eeq
If the reaction is to happen at all, then the RHS of
(\ref{eq:pbot}) must be positive.
The points 
where $p'_\bot=0$ give the upper and lower
bounds $p'_{z1},p'_{z2}$ for the integral (\ref{eq:gendedt}).
Well above threshold the mass term becomes negligible,
and the longitudinal phase space is determined by
those $z$ values for which $\pm \xi - \eta_+(2z-1)$ is positive
in the interval  $z\in[0,1]$.


Using these results in the expression (\ref{eq:gendedt}) for the
decay rate we find that, well above threshold, the rate for decay
of a positive helicity photon to a positive helicity electron and
negative helicity positron is given by
\beq \Gamma\approx \frac{e^2k^2}{48\pi M} \left[\left(2\xi +
\eta_+\right)\, \Theta(\xi + \eta_+) + 
\frac{(\xi-\eta_+)^4}{16|\eta_+|^3}\, \Theta(|\eta_+|-|\xi|) \right]. \eeq
where $\Theta(x)$ is the Heaviside step function.
This agrees up to a numerical factor with the estimate
in~\cite{Gelmini:2005gy} in the case where $\eta_+=0$.

For the purposes of imposing constraints, it is important to
know how rapidly the decay rate becomes large above threshold.
We will return to this question in section~\ref{subsec:phot-dec-ratenear}, after we derive the photon decay threshold.

\subsection{Photon decay threshold}
\label{subsec:phot-dec-thr}

The photon decay threshold is found by setting $p^{'2}_\bot=0$ and
solving for the minimum value of $k$ for which there is a solution
for some $z$. From the threshold analysis described in
section~\ref{apsec:threshconfig} we know that the outgoing momenta
are parallel, hence $0<z<1$ at threshold, so the minimum must have
$z$ within this range. 

Let us specialize to the case of a positive helicity photon (the negative helicity case works similarly).  In terms of the variable $x=1-2z$, the
condition that (\ref{eq:pbot}) vanish then becomes 
\beq \label{eq:pbot3} (1-x^2)(\xi + \eta_+x ) = \frac {4m^2M} {k^3}.
\eeq
The range of $x$ is $-1<x<1$, so we see that if $|\eta_+|$ is
sufficiently large for a given $\xi$ the conservation equation has a
solution.  Hence the threshold depends on $\eta_+$ only through its
absolute value (the equation is invariant under $\eta_+, x
\rightarrow -\eta_+,-x$). Physically, this corresponds to the fact
that since the positron and electron have opposite dispersion, one
is always subluminal. By depositing most of the outgoing momentum in
the subluminal particle we can magnify its LV while minimizing the
LV of the superluminal partner.  Hence for either sign of $\eta_+$
we can find a solution.

Since the RHS of (\ref{eq:pbot3}) decreases as $k$ increases, the
threshold will occur when the LHS is at its maximum in $x$. Let us
consider a few special cases. First, if If $\eta_+=0$, then the LHS
has a positive maximum only if $\xi>0$. This occurs at $x=0$ (equal
electron and positron momenta) and is equal to $\xi$. In this case
the threshold $k_{th}$ satisfies $\xi=4m^2M/k_{th}^3$. If instead
$\xi=0$, then the maximum of the LHS occurs at $x=1/\sqrt{3}$, and
is equal to $2|\eta_+|/3\sqrt{3}$. In this case the threshold
$k_{th}$ satisfies $|\eta_+|=6\sqrt{3}m^2M/k_{th}^3$. If
$\xi<-|\eta_+|$, the LHS is never positive in the range $-1<x<1$, so
photon decay does not occur for such parameters. More generally, the
maximum of the LHS occurs at
\beq \label{eq:x} x_{max}= \frac{-\xi +
\sqrt{\xi^2+3\eta_+^2}}{3\eta_+}. \eeq
Substituting $x_{max}$ into (\ref{eq:pbot3}) allows us to find the
maximum of the LHS, but the expression is a bit complicated so we
don't bother to display it here. One can use it to solve for $\xi$
as a function of $\eta_+$ and $k_{th}$ and thus derive a constraint.

\subsection{Photon decay rate near threshold}
\label{subsec:phot-dec-ratenear}

We now return to the question of reaction rate above threshold. If
particle lifetimes are short slightly above threshold, then we are
justified in using the absence of a threshold to establish
constraints on the parameters. As a simple example of the rapidity
with which reaction rates increase above threshold, let us analyze
photon decay with $\xi=0$, a positive helicity incoming photon, and
$\eta_+>0$ (other parameter choices and the Cerenkov effect work
similarly).

We expand the initial photon momentum as $k=(1+\alpha) k_{th}$,
where $k_{th}^3=6\sqrt{3}m^2M/|\eta_+|$ and $\alpha \ll 1$.  At
threshold the entire phase space consists of one point,
$x=x_{max}=1/\sqrt{3}$, and $p'_\bot$ vanishes at this point. To
calculate the rate using Eqns.~\ref{eq:gendedt}
and~\ref{eq:phot-dec-M} we need to know the range in $x$ slightly
above threshold and the value of $p'_\bot$ in this range to first
order in $\alpha$.  The rate is then given by
\begin{equation} \label{eq:phot-decay1}
\Gamma(p) =  \frac {e^2} {16 \pi k} \int_{x_{max}-\Delta
x}^{x_{max}+\Delta x}  dx \frac {1+x} {1-x} p_\bot^{'2}(x),
\end{equation}
where we have rewritten everything in terms of $x$.  $\Delta x$ is
the spread around $x_{max}$ as we move above threshold.  The fact
that $x_{max}$ is an extremum implies that spread around $x_{max}$
is symmetric, and just above threshold the spread is small, $\Delta
x\ll x_{max}$.  Expanding $p_\bot^{'2}$ about $x_{max}$ as a
parabola, we find Eq.~\ref{eq:phot-decay1} is approximately given by
\begin{equation} \label{eq:rateabove}
\Gamma(p)=\frac {e^2} {12 \pi k_{th}} \frac {1+x_{max}} {1-x_{max}}
p_\bot^{'2}(x_{max})\Delta x.
\end{equation}

Now $p_\bot^{'2} (x_{max})$ is given by Eq.~\ref{eq:pbot} at
$k=k_{th}(1+\alpha)$, i.e.
\beq   p_\bot^{'2}(x_{max})=\eta_+ x_{max} \frac {1-x_{max}^2} {4}
k_{th}^3 (1+\alpha)^3 - m^2. \eeq
{}From the definition of $k_{th}$ we know that the zeroth order term
in $\alpha$ vanishes.  With this fact we can easily solve for
$p_\bot^{'2}(x_{max})=3 \alpha m^2$ (to first order in $\alpha$).

All that remains to evaluate is $\Delta x$.  At the endpoints of
integration in Eq. \ref{eq:gendedt} we know that $p'_\bot=0$, which
implies that we have the relation
\begin{equation} \label{eq:deltax1}
\eta_+ (x_{max}\pm \Delta x) \frac {1-(x_{max}\pm\Delta x)^2} {4}
k_{th}^3 (1+\alpha)^3 - m^2 = 0.
\end{equation}
Expanding Eq.~\ref{eq:deltax1} to lowest order in $\Delta x$ and
$\alpha$ yields
\begin{equation}
x_{max}(1-x_{max}^2)+3 \alpha x_{max} (1-x_{max}^2) - 3 x_{max}
(\Delta x)^2= \frac {4 m^2} {k_{th}^3 \eta_+}.
\end{equation}
There is no linear term in $\Delta x$ since $x_{max}$ is an extremum
of Eq.~\ref{eq:deltax1}. The first term on the left hand side is
equal to the right hand side by definition of $k_{th}$ so we obtain
$(\Delta x)^2=2\alpha/3$. Substituting $\Delta x$ and
$p_\bot^{'2}(x_{max})$ into Eq.~\ref{eq:rateabove} yields our final
expression for the rate as a function of $\alpha$,
\begin{equation}
 \Gamma= \frac{e^2m^2}{4\pi k_{th}}\sqrt{2/3}(2 +
\sqrt{3})\alpha^{3/2}
\end{equation}

To see how quickly the reaction starts to happen above threshold,
consider an incoming photon of energy 1\% above threshold. The
lifetime of the photon is equal to $10^{-8} k_{th}/10~$TeV seconds,
short enough that on any relevant astrophysical timescale a
population of photons of multi-TeV energies will almost completely
decay. As this example demonstrates, since the lifetime is extremely
short only slightly above threshold, we are justified in using
threshold values to derive constraints.

\subsection{Vacuum \v{C}erenkov threshold and rate}
\label{sebsec:cheren-thr}

The vacuum \v{C}erenkov process $e\rightarrow e\gamma$ is forbidden
by angular momentum conservation in the threshold configuration.
However, as with photon decay to opposite helicities, the rate
becomes large above threshold. In fact the rate calculation is very
close to that for photon decay.\footnote{An extensive investigation of
vacuum \v{C}erenkov radiation was
carried out in Ref.~\cite{Lehnert:2004hq}, focusing on
the case of a Maxwell-Chern-Simons, dimension three LV operator for the photon.
Unfortunately, that form of LV is different from what we study here, hence we cannot
make direct use of those results.}
 The only significant difference is
that a factor of the outgoing photon energy $\omega(k)$ must be
inserted into (\ref{eq:gendedt}) since we are concerned with the
energy loss rate.  The net result is that $dE/dt \sim p^3/M$ above
threshold, which implies that a 10 TeV electron would emit a
significant fraction of its energy in $10^{-9}$ seconds.  Hence in
this case as well a threshold analysis will give accurate results.

We now summarize the results of the threshold analysis. To begin
with, if the photon dispersion is unmodified and the electron
parameter $\eta$ (for one helicity) is positive, then the electron
group velocity $v_g=1-(m^2/2p^2)+(\eta p/M)+\cdots$ exceeds the
speed of light when
\beq p_{\rm th} =(m^2M/2\eta)^{1/3}\simeq 11\, {\rm TeV}\,
\eta^{-1/3}. \label{pcerenkov} \eeq
This turns out to be the threshold energy for the vacuum
\v{C}erenkov process with emission of a zero energy photon, which
we call the soft \v{C}erenkov threshold. There is also the
possibility of a hard \v{C}erenkov
threshold\cite{Jacobson:2002hd,KonMaj}. For example, if the
electron dispersion is unmodified and the photon parameter $\xi$
is negative then at sufficiently high electron energy the emission
of an energetic positive helicity photon is possible. This hard
\v{C}erenkov threshold occurs at $p_{\rm th}=(-4m^2M/\xi)^{1/3}$,
and the emitted photon carries away half the incoming electron
momentum. It turns out that the threshold is soft when both
$\eta>0$ and $\xi\ge -3\eta$, while it is hard when both $\xi
<-3\eta$ and $\xi<\eta$. The hard threshold in the general case is
given by $p_{\rm th}=(-4m^2M(\xi+\eta)/(\xi-\eta)^2)^{1/3}$, and
the photon carries away a fraction $(\xi-\eta)/2(\xi+\eta)$ of the
incoming momentum. In the general case at threshold, neither the
incoming nor outgoing electron group velocity is equal to the
photon phase or group velocity, so the hard \v{C}erenkov effect
cannot simply be interpreted as being due to faster than light
motion of a charged particle.

\subsection{Fermion pair emission threshold}
\label{app:pairemission}

The threshold for the
process $e^- \rightarrow e^-e^-e^+$ involves only one LV parameter
when the electrons all have the same helicity and the positron has the opposite
helicity. In the threshold configuration this is allowed by angular momentum
conservation, and the form of the
spinors (\ref{spinorshighE},\ref{pspinorshighE}) shows that the amplitude is
large as well, so rate will be high above threshold. It thus suffices to
determine the threshold.

Suppose the electron has positive helicity, and suppose that
$\eta_+>0$, so the reaction will proceed in this configuration.
Let the incoming electron have momentum $p$, and let
the positron have momentum $zp$.
Since the electron dispersion relation then has positive curvature
everywhere, as discussed in Appendix \ref{apsec:threshconfig}
the two final state electrons will
have the same momentum, $(1-z)p/2$. Writing out the conservation of
energy in the threshold configuration, and keeping as usual just
the first order terms in $m^2$ and $\eta_+$, one finds the relation
\begin{equation} \label{eq:pairemission}
p^3= \frac{2 m^2M} {\eta_+z(1-z)}
\end{equation}
The right hand side is minimized when $z=1/2$, yielding the
threshold formula
\beq p_{\rm th} =(8m^2M/\eta)^{1/3},\label{ppairemission} \eeq
which is a factor $16^{1/3}\sim 2.5$ above the soft \v{Cerenkov}
theshold (\ref{pcerenkov}).

\subsection{Helicity decay rate}
\label{subsec:HelDec}

A variation on the \v{Cerenkov} effect that has received almost no
attention in the literature is ``helicity decay". If $\eta_+$ and $
\eta_-$ are unequal, say $\eta_->\eta_+$, then a negative helicity
electron can decay into a positive helicity electron and a photon,
even when the LV parameters do not permit the vacuum \v{C}erenkov
effect. In this process, the large $R$ or small ($O(m/E)$) $L$
component of a positive helicity electron is coupled to the small
$R$ or large $L$ component of a negative helicity electron
respectively.\footnote{Note that the relative size of $L$ and $R$
components of the fermion spinors is still controlled by the mass,
as can be seen from Eq.~\ref{eq:diraceqn}.}  Such helicity decay has
no threshold energy, so whether this process can be used to set a
constraint is solely a matter of the decay rate. As we shall see
however, the rate is maximum at an ``effective threshold" energy
below which it is very strongly suppressed, and above which it
decreases as $1/E$. Hence in practice there is a threshold for
helicity decay as well.

The helicity decay rate involves all three LV parameters $\eta_\pm$
and $\xi$. However, since the birefringence constraints limit $\xi$
to a level far beyond where helicity decay is sensitive, we set
$\xi=0$ for the purposes of evaluating the helicity decay rate here.

We specialize to the case of a negative helicity electron decaying
into a positive helicity electron, the opposite case works
identically. Again, the first step is to calculate the matrix
element for the reaction, given by
\begin{equation} \label{eq:MEhd}
{\rm i}{\sf M}={\rm i}e \bar{u}_{+}(p') \epsilon^{\pm *}_\alpha
\gamma^\alpha u_-(p)
\end{equation}
where $p,p'$ are the incoming and outgoing electron momenta,
respectively. The calculation proceeds along the lines of photon
decay, and the matrix element can be evaluated approximately as
\begin{equation} \label{eq:helicitydecayM}
{\rm i}{\sf M}={\rm i}e m \left( \sqrt{\frac {E} {E'}} - \sqrt{\frac
{E'} {E}} \right).
\end{equation}
No angular dependence occurs in this lowest order approximation for
{\sf M} since the helicity flip allows angular momentum to be
conserved when all momenta are parallel.

There are two different regimes for the size of the matrix element
and rate, separated by the momentum $p_{th}\approx
(m^2M/(\eta_--\eta_+))^{1/3}$. To see this, we must turn to the
conservation equation.  We will be interested in the amount of
longitudinal phase space available to the reaction, i.e. the bounds
$p'_{z1}$ and $p'_{z2}$ in Eq.~\ref{eq:gendedt}. The energy
conservation equation without any transverse momenta is
\begin{equation}\label{eq:helicitycons0}
2p+\frac{m^2} {p} + \eta_- \frac{p^2}{M}= 2|p'|+\frac {m^2} {|p'|} +
\eta_+ \frac{p'^2}{M}+ 2|k|
\end{equation}
where $p'$ and $k$ are the longitudinal components of the final
electron and photon momenta. We introduce the variable $z$ by $k=pz,
p'=p(1-z)$. The photon energy is $|k|$, so energy conservation
implies $|z|<1$. Since this is not a threshold situation, the final
momenta can be anti-parallel, so negative values of $z$ are
permitted. In terms of $z$, Eq.~\ref{eq:helicitycons0} becomes
\begin{equation} \label{eq:helicitycons1}
\frac{m^2} {p^2} = \bigg{[} \frac {p} {M} (\eta_- - \eta_+(1-z)^2) -
2(|z|-z) \bigg{]} \,\frac {1-z} {z}.
\end{equation}

If $z<0$ then the lowest order solution (using $m^2/p^2 \ll 1$ and
$p/M \ll 1$) is $z=-(\eta_- - \eta_+) p (4M)^{-1}$. This is negative
only if $\eta_- > \eta_+$ and is a solution for any $p$. Hence,
there is no absolute threshold.  In practice, $p/M$ is so small that
we can neglect the difference between this value of $z$ and zero. We
shall therefore set the lower bound on our rate integral for
helicity decay to zero.

Now consider $z>0$.  Eq. \ref{eq:helicitycons1} can in this case be
rewritten as
\begin{equation} \label{eq:helicitycons2}
\frac{m^2M} {p^3} = \bigg{[} \eta_- - \eta_+(1-z)^2 \bigg{]}
\,\frac {1-z} {z}.
\end{equation}
At low energies ($m^2M/p^3 \gg (\eta_--\eta_+)$), the value of $z$
that solves Eq. \ref{eq:helicitycons2} is also very small, $z
\approx (\eta_--\eta_+) p^3/(m^2M)=\bar{z}$.  The corresponding
bounds on the integral (\ref{eq:gendedt}) are therefore zero and
$\bar{z}$, so there is only a very small amount of phase space
available for the reaction. Furthermore, since $z$ is small, $E'
\approx E$ and (\ref{eq:helicitydecayM}) is doubly suppressed, once
by the mass and once by $\Delta E=E-E'$. Evaluating $M$ explicitly
at $z=\bar{z}$ gives
\begin{equation} \label{eq:helicityMlowx}
{\rm i}{\sf M} \approx {\rm i}e (\eta_--\eta_+) \frac {p^3} {mM}.
\end{equation}
We can overestimate the rate at low energies by substituting
(\ref{eq:helicityMlowx}) and the bounds on $z$ into
(\ref{eq:gendedt}). It is an overestimate since at $z=\bar{z}$ the
photon has its maximum possible energy, so the difference between
$E$ and $E'$ is maximized, hence we are using the largest possible
value of {\sf M} in the allowed region of phase space. Upon
integration the decay rate is {\it at most}
\begin{equation}
\Gamma_1=\frac {e^2} {16 \pi} \frac {(\eta_--\eta_+)^3 p^8} {m^4
M^3}. \label{eq:lowrate}
\end{equation}

If $p\gg (m^2M/(\eta_--\eta_+))^{1/3}$ then the situation is much
different. The solution to the conservation equation
(\ref{eq:helicitycons2}) has $z\approx 1$ so the entire longitudinal
phase space has opened up. Therefore $E' \neq E$ and the only
suppression is by the mass. The resulting decay rate in this case is
approximately
\begin{equation}
\Gamma_2=\frac{e^2 m^2} {16 \pi p} \times O(1), \label{eq:highrate}
\end{equation}
which roughly coincides with $\Gamma_1$ at the transition momentum
$p_{tr}= (m^2M/(\eta_--\eta_+))^{1/3}$ (as it must).  Note that the
decay rate actually decreases with increasing momentum.  This is
because the electrons become more and more chiral as momentum
increases, thereby reducing their ability to flip helicity via the
QED vertex.

For $\eta_{-}-\eta_+$ of O(1), $p_{tr}$ is approximately 10 TeV.
Hence a 1 TeV negative helicity electron has a decay rate given by
$\Gamma_1$ which yields a lifetime of about one second.  In contrast
a 50 TeV electron with a decay rate of $\Gamma_2$ has a lifetime of
approximately $10^{-9}$ seconds, making a simple ``threshold''
analysis applicable for deriving constraints if 50 TeV negative
helicity electrons are required to exist for the longer time scales
relevant in a system such as the Crab nebula.

\section*{Acknowledgements}

This research was supported in part by the NSF under grants
PHY-9800967 and PHY-0300710 at the University of Maryland and by
the DOE under grant DE-F603-91ER40674 at UC Davis.  DM wishes to
thank the Mitchell Institute for hospitality during some of this
work. SL wishes to thank M.~Colpi, S.~Cristiani, M.~Magliocchetti
and E.~Pian for stimulating discussions about astrophysical
observations and models. The authors wish to thank A.~A.~Andrianov
for pointing out the omission of 
fermion pair emission in the
first draft of this work.





\end{document}